\documentclass{article}

\usepackage{arxiv}

\usepackage[utf8]{inputenc} 
\usepackage[T1]{fontenc}    
\usepackage{lmodern}        
\usepackage{hyperref}       
\usepackage{url}            
\usepackage{booktabs}       
\usepackage{amsfonts}       
\usepackage{nicefrac}       
\usepackage{microtype}      
\usepackage{graphicx}

\title{Fine-resolution landscape-scale biomass mapping using a spatiotemporal patchwork of LiDAR coverages}

\author{
    Lucas K Johnson
   \\
    Graduate Program in Environmental Science \\
    State University of New York College of Environmental Science and Forestry \\
  Syracuse, NY 13210 \\
  \texttt{\href{mailto:ljohns11@esf.edu}{\nolinkurl{ljohns11@esf.edu}}} \\
   \And
    Michael J Mahoney
   \\
    Graduate Program in Environmental Science \\
    State University of New York College of Environmental Science and Forestry \\
  Syracuse, NY 13210 \\
  \texttt{\href{mailto:mjmahone@esf.edu}{\nolinkurl{mjmahone@esf.edu}}} \\
   \And
    Eddie Bevilacqua
   \\
    Department of Sustainable Resources Management \\
    State University of New York College of Environmental Science and Forestry \\
  Syracuse, NY 13210 \\
  \texttt{\href{mailto:ebevilacqua@esf.edu}{\nolinkurl{ebevilacqua@esf.edu}}} \\
   \And
    Stephen V Stehman
   \\
    Department of Sustainable Resources Management \\
    State University of New York College of Environmental Science and Forestry \\
  Syracuse, NY 13210 \\
  \texttt{\href{mailto:svstehma@esf.edu}{\nolinkurl{svstehma@esf.edu}}} \\
   \And
    Grant Domke
   \\
    Northern Research Station \\
    USDA Forest Service \\
  St.~Paul, MN 55114 \\
  \texttt{\href{mailto:grant.m.domke@usda.gov}{\nolinkurl{grant.m.domke@usda.gov}}} \\
   \And
    Colin M Beier
   \\
    Department of Sustainable Resources Management \\
    State University of New York College of Environmental Science and Forestry \\
  Syracuse, NY 13210 \\
  \texttt{\href{mailto:cbeier@esf.edu}{\nolinkurl{cbeier@esf.edu}}} \\
  }

\usepackage{longtable,booktabs,array}
\usepackage{calc} 
\usepackage{etoolbox}
\makeatletter
\patchcmd\longtable{\par}{\if@noskipsec\mbox{}\fi\par}{}{}
\makeatother
\IfFileExists{footnotehyper.sty}{\usepackage{footnotehyper}}{\usepackage{footnote}}
\makesavenoteenv{longtable}

\newlength{\cslhangindent}
\newlength{\csllabelwidth}
\newlength{\cslentryspacingunit} 
  {}%
  {\par}
\newenvironment{CSLReferences}[2] 
 {
  \setlength{\parindent}{0pt}
  \ifodd #1
  \let\oldpar\par
  \def\par{\hangindent=\cslhangindent\oldpar}
  \fi
  \setlength{\parskip}{#2\cslentryspacingunit}
 }%
 {}
\usepackage{calc}

\usepackage{amsmath}
\usepackage{booktabs}
\usepackage{longtable}
\usepackage{array}
\usepackage{multirow}
\usepackage{wrapfig}
\usepackage{float}
\usepackage{colortbl}
\usepackage{pdflscape}
\usepackage{tabu}
\usepackage{threeparttable}
\usepackage{threeparttablex}
\usepackage[normalem]{ulem}
\usepackage{makecell}
\usepackage{xcolor}
\begin{document}
\maketitle

\begin{abstract}
Estimating forest aboveground biomass (AGB) at large scales and fine spatial
resolutions has become increasingly important for greenhouse gas accounting,
monitoring, and verification efforts to mitigate climate change.
Airborne LiDAR is highly valuable for modeling attributes of forest structure
including AGB,
yet most LiDAR collections take place at local or regional scales
covering irregular, non-contiguous footprints,
resulting in a patchwork of different landscape segments at various points
in time.
Here,
as part of a statewide forest carbon assessment for New York State (USA),
we addressed common obstacles in leveraging a LiDAR patchwork for AGB mapping
at landscape scales, including selection of training data,
the investigation of regional or coverage specific patterns in prediction
error,
and map agreement with field inventory across multiple scales.

Three machine learning algorithms and an ensemble model were trained with
Forest Inventory and Analysis (FIA) field measurements,
airborne LiDAR, and topographic, climatic, and cadastral geodata.
Using a novel set of plot selection criteria and growth adjustments
to temporally align LiDAR coverages with FIA measurements,
801 FIA plots were selected with co-located point clouds drawn from a
patchwork of 17 leaf-off LiDAR coverages (2014-2019).
Our ensemble model was used to produce 30 m AGB prediction surfaces within a
predictor-defined area of applicability (98\% of LiDAR coverage)
and within four vegetated landcover classes.
The resulting AGB maps were compared with FIA
plot-level and areal estimates at multiple scales of aggregation.
Our model was overall accurate (\% root mean squared error 22-45\%;
mean absolute error 11.6-29.4 Mg ha\textsuperscript{-1}; mean error 2.4-6.3 Mg ha\textsuperscript{-1}),
explained 73-80\% of field-observed variation,
and yielded estimates that were largely consistent with FIA's design-based
estimates (89\% of estimates within FIA's 95\% confidence interval).
We share practical solutions to challenges faced in using spatiotemporal
patchworks of LiDAR to meet growing needs for AGB prediction and mapping
in support of broad-scale applications in forest carbon accounting and
ecosystem stewardship.

\,
\end{abstract}


\hypertarget{introduction}{%
\section{Introduction}\label{introduction}}

Mapping and monitoring forest aboveground biomass (AGB) has become increasingly
important as the basis for large-scale accounting of carbon and greenhouse gas
(GHG) fluxes in support of policy, regulatory, and land stewardship initiatives
to mitigate global climate change.
Although carbon stocks are the ultimate endpoint in GHG accounting,
for several practical reasons forest AGB serves as a proxy for carbon stocks
(and stock-changes) in large-scale accounting methodologies
(Buendia et al. 2019; Woodall et al. 2015).
Such applications,
based primarily on field inventory,
yield aggregate tabular estimates for political units (e.g., states, provinces),
but offer little insights on landscape patterns within and across those units.
Landscape-scale AGB maps with fine-resolutions can serve as inputs to GHG
accounting
and can help decision makers identify specific units of land for protection from
deforestation, or as suitable candidates for reforestation,
afforestation, or improved management in efforts to increase terrestrial carbon
sequestration and offset GHG emissions from other sectors
(R. A. Houghton et al. 2012; R. Houghton 2005).

National field sampling~programs,
like the United States Department of Agriculture's Forest Inventory and Analysis
program (FIA) (Gray et al. 2012),
provide estimates of AGB at individual plots and are scaled to larger areas
through a design-based approach (Bechtold and Patterson 2005).
However, the resolution at which design-based estimates can be reliably
produced is limited to relatively coarse units, such as states or counties,
due to the density of the sample (McRoberts 2011).
For instance, the FIA program samples forest conditions and land cover across
New York State (NYS) at a density of roughly one plot per
2,400 ha (Gray et al. 2012),
while 65\% of NY forestlands are owned in parcels smaller than
40 ha
(L'Roe and Allred 2013).
Although FIA's extensive plot network was never designed to yield such
localized results,
this inherent resolution mismatch poses obstacles to accounting and
decision-support applications that must account for local geography,
both ecological and cadastral, to be practical and effective.

To address this need,
model-based approaches combining field inventory data,
like the FIA,
with auxiliary remotely sensed data can produce predictions for all map units
(pixels) in a given area.
Airborne LiDAR has been established as a highly valuable remotely sensed data
source for such purposes (Huang et al. 2019; Hurtt et al. 2019; Chen and McRoberts 2016) offering
detailed information on forest structure at fine spatial resolutions.
However, LiDAR data are most commonly acquired at local to regional scales in
irregular or non-contiguous footprints (Skowronski and Lister 2012),
resulting in a complex patchwork of data from component coverages acquired
at different times with different sensors and mission parameters,
in turn posing a host of challenges for broad-scale AGB mapping
(Lu et al. 2014; Huang et al. 2019).
These challenges include insufficient field inventory plots that spatially and
temporally match LiDAR acquisitions and data discrepancies among LiDAR
coverages.

Yet several groups have undertaken broad-scale AGB
mapping efforts with LiDAR patchworks
with varying degrees to which training data have been pooled
from multiple LiDAR coverages.
The choice between an individual or a pooled modeling approach often reflects
practical considerations relating to sufficient sample size across all
sub-regions and the cost of developing multiple models.
Nilsson et al. (2017) did not pool at all,
implementing a separate model trained for each coverage.
Huang et al. (2019) pooled by ecoregion.
Both Ayrey et al. (2021) and Hauglin et al. (2021) pooled all coverages but used a
convolutional neural network and a mixed-effects model respectively, with
differing protocols for inventory plot selection.

In this study,
as part of a broader effort for map-based forest carbon accounting across NYS,
we addressed several common challenges in using LiDAR
patchworks for broad-scale, fine-resolution biomass modeling and mapping.
We leveraged FIA inventories for model training and assessment data,
and implemented FIA-developed methods to assess the agreement between our
estimates and those produced by FIA.
With the goal of producing a spatially explicit representation of FIA AGB
information,
we used a model-based approach to translate FIA's discrete plot-level estimates
to wall-to-wall predictions at a 30 m resolution across a patchwork of 17
discrete LiDAR coverages in NYS.

We implemented a rigorous plot selection framework to limit temporal lags
between LiDAR acquisitions and field inventories.
When strict temporal alignment yielded too few plots,
we leveraged repeated FIA inventories to boost the number of plots that
temporally match LiDAR acquisitions without the incorporation of additional
models or manual processes (Ayrey et al. 2021; Hauglin et al. 2021).
In spite of this strategy, we were left with limited model training data, where
some coverages and regions lacked sufficient information to support independent
models, necessitating a single model or pooled approach.

We used machine learning (ML) algorithms including
random forests (Breiman 2001a), gradient boosting machines (Friedman 2002),
and support vector machines (Cortes and Vapnik 1995), as well as `stacked ensembles' of
said algorithms (Wolpert 1992).
These have been shown to be better suited for pure prediction
(Efron 2020; Breiman 2001b)
when compared to their conventional regression counterparts (Hauglin et al. 2021),
especially when the input data are noisy and the processes governing the
relationships between predictors and responses are complex or unknown, as
is often the case in nature (Wintle et al. 2003).
Furthermore, using pre-selected hyperparameters these algorithms can be trained
in minutes with typical consumer-grade hardware, avoiding some of the
drawbacks of computationally intense deep-learning approaches (Ayrey et al. 2021).

We also employed multiple strategies to address concerns that differences among
sensors and mission parameters could lead to non-randomly distributed errors
in our AGB prediction surfaces.
First, we produced an area of applicability (Meyer and Pebesma 2021) mask to both
examine the uniformity of our predictors across the component coverages,
as well as to screen predictions based on anomalous predictor data.
Second, we examined the spatial autocorrelation of our residuals and mapped
our prediction error to identify the presence of region or coverage specific
patterns.
Finally, we assessed the agreement between our mapped predictions and FIA
estimates across a range of scales (Riemann et al. 2010; Menlove and Healey 2020).

\hypertarget{data-and-methods}{%
\section{Data and Methods}\label{data-and-methods}}

\hypertarget{lidar-coverages}{%
\subsection{LiDAR Coverages}\label{lidar-coverages}}

\begin{figure}
\includegraphics[width=1\linewidth]{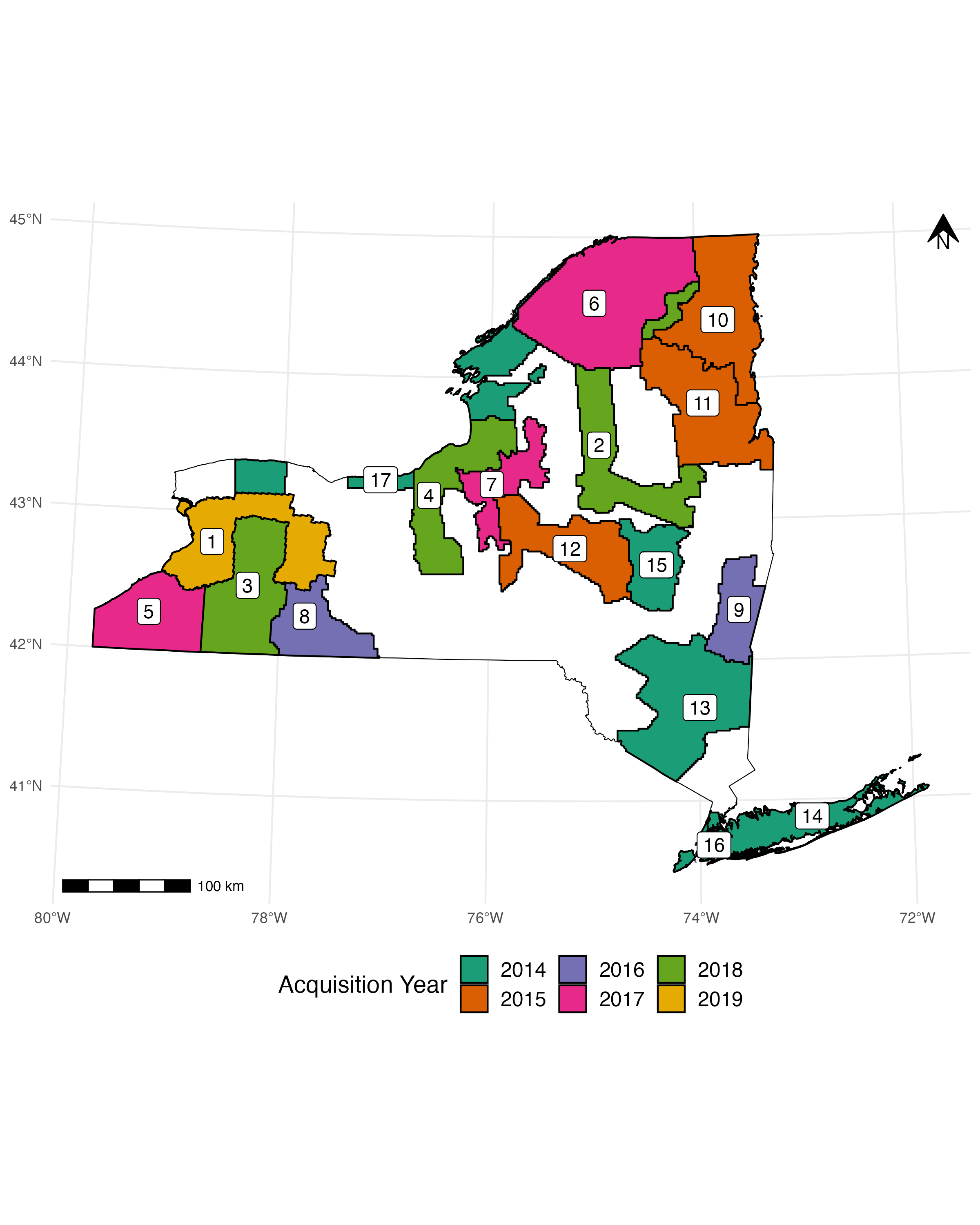} \caption{GPO-LiDAR component coverages, colored by year of acquisition and labeled by ID numbers in Table 1.}\label{fig:regions}
\end{figure}

Our study relied upon a set of
17
LiDAR datasets hosted by the NYS GIS Program Office (GPO) covering
62.46\%
(7,835,690
ha) of NYS (hereafter ``GPO-LiDAR area''; Table \ref{tab:lidarsummary};
Figure \ref{fig:regions}).
We selected individual coverages from the most recent five years of available
data that contained temporally matching field data (2014-2019) to minimize
sensor and data differences.
All component coverages were originally collected to generate digital elevation
models for flood risk analysis, and to this end were flown during leaf-off
conditions.
Several previous studies have shown that leaf-off LiDAR models can be as
accurate as their leaf-on counterparts
(Hawbaker et al. 2010; White et al. 2015; Anderson and Bolstad 2013).

\begin{table}

\caption{\label{tab:lidarsummary}Component LiDAR coverage metadata. IDs for cross figure correspondence; Year of acquisition; Area covered (ha); Pulse density (PD) in pulses per m$^2$; Number of FIA plots in model and map assessment datasets. Area of Applicability (AOA) in percent of LiDAR coverage pixels considered inside of the area of applicability. AOA computation conducted after initial LCMAP masking.}
\centering
\fontsize{10}{12}\selectfont
\begin{tabular}[t]{lllrrrrr}
\toprule
\multicolumn{1}{c}{ID} & \multicolumn{1}{c}{Coverage Name} & \multicolumn{1}{c}{Year} & \multicolumn{1}{c}{Area} & \multicolumn{1}{c}{PD} & \multicolumn{1}{c}{\% AOA} & \multicolumn{1}{c}{Model} & \multicolumn{1}{c}{Assessment}\\
\midrule
1 & Erie, Genesee \& Livingston & 2019 & 555,853 & 2.04 & 97.72 & 11 & 33\\
\addlinespace
2 & Fulton, Saratoga, Herkimer \& Franklin & 2018 & 557,421 & 2.60 & 99.61 & 39 & 84\\
\addlinespace
3 & Southwest B & 2018 & 527,075 & 1.98 & 98.06 & 28 & 66\\
\addlinespace
4 & Cayuga \& Oswego & 2018 & 438,201 & 2.78 & 96.89 & 25 & 43\\
\addlinespace
5 & Southwest & 2017 & 423,714 & 1.98 & 98.53 & 33 & 74\\
\addlinespace
6 & Franklin \& St. Lawrence & 2017 & 977,620 & 2.69 & 99.24 & 104 & 188\\
\addlinespace
7 & Oneida Subbasin & 2017 & 264,886 & 2.10 & 96.02 & 15 & 31\\
\addlinespace
8 & Allegany \& Steuben & 2016 & 309,081 & 1.69 & 97.77 & 32 & 53\\
\addlinespace
9 & Columbia \& Rensselaer & 2016 & 248,839 & 1.69 & 98.19 & 17 & 28\\
\addlinespace
10 & Clinton, Essex \& Franklin & 2015 & 600,755 & 2.23 & 98.70 & 115 & 127\\
\addlinespace
11 & Warren, Washington \& Essex & 2015 & 611,704 & 3.24 & 99.37 & 106 & 128\\
\addlinespace
12 & Madison \& Otsego & 2015 & 471,564 & 2.13 & 99.24 & 56 & 92\\
\addlinespace
13 & 3 County & 2014 & 755,629 & 2.04 & 96.12 & 92 & 114\\
\addlinespace
14 & Long Island & 2014 & 315,542 & 2.04 & 94.77 & 22 & 12\\
\addlinespace
15 & Schoharie & 2014 & 256,464 & 2.04 & 95.12 & 31 & 40\\
\addlinespace
16 & New York City (NYC) & 2014 & 77,211 & 1.54 & 90.40 & 2 & 1\\
\addlinespace
17 & Great Lakes & 2014 & 444,215 & 2.04 & 98.47 & 73 & 103\\
\midrule
\addlinespace
\textbf{} & \textbf{GPO-LiDAR} & \textbf{} & \textbf{7,835,773} & \textbf{} & \textbf{98.12} & \textbf{801} & \textbf{1,217}\\
\bottomrule
\end{tabular}
\end{table}

\hypertarget{field-data}{%
\subsection{Field Data}\label{field-data}}

Two field datasets were compiled from the inventory USDA FIA inventory
in NYS with the distinct purposes of model development and map assessment.
The FIA program compiled AGB estimates for trees \(\geq\) 12.7 cm (5 in) diameter
at breast height (Gray et al. 2012),
and were converted to units of megagrams per hectare (Mg ha\textsuperscript{-1}).
The FIA uses permanent inventory plots arranged
in a quasi-systematic hexagonal grid that are
remeasured on a rolling seven-year basis in the Northeastern United States
(Bechtold and Patterson 2005).
Tree measurements, and subsequently AGB estimates, are only recorded on portions
of plots considered forested.
For an area to be considered forested by the FIA, the area must be at least
10\% stocked with trees, at least
0.4 ha
(1 acre) in size, and at least
36.58 m (120 ft) wide.
Additionally, any lands meeting these minimum requirements,
but developed for nonforest land uses,
are not considered forested.
With an understanding that some nonforest conditions contained AGB that was
not measured by FIA, we assumed that any nonforest condition contained 0 AGB.

\begin{figure}
\includegraphics[width=1\linewidth]{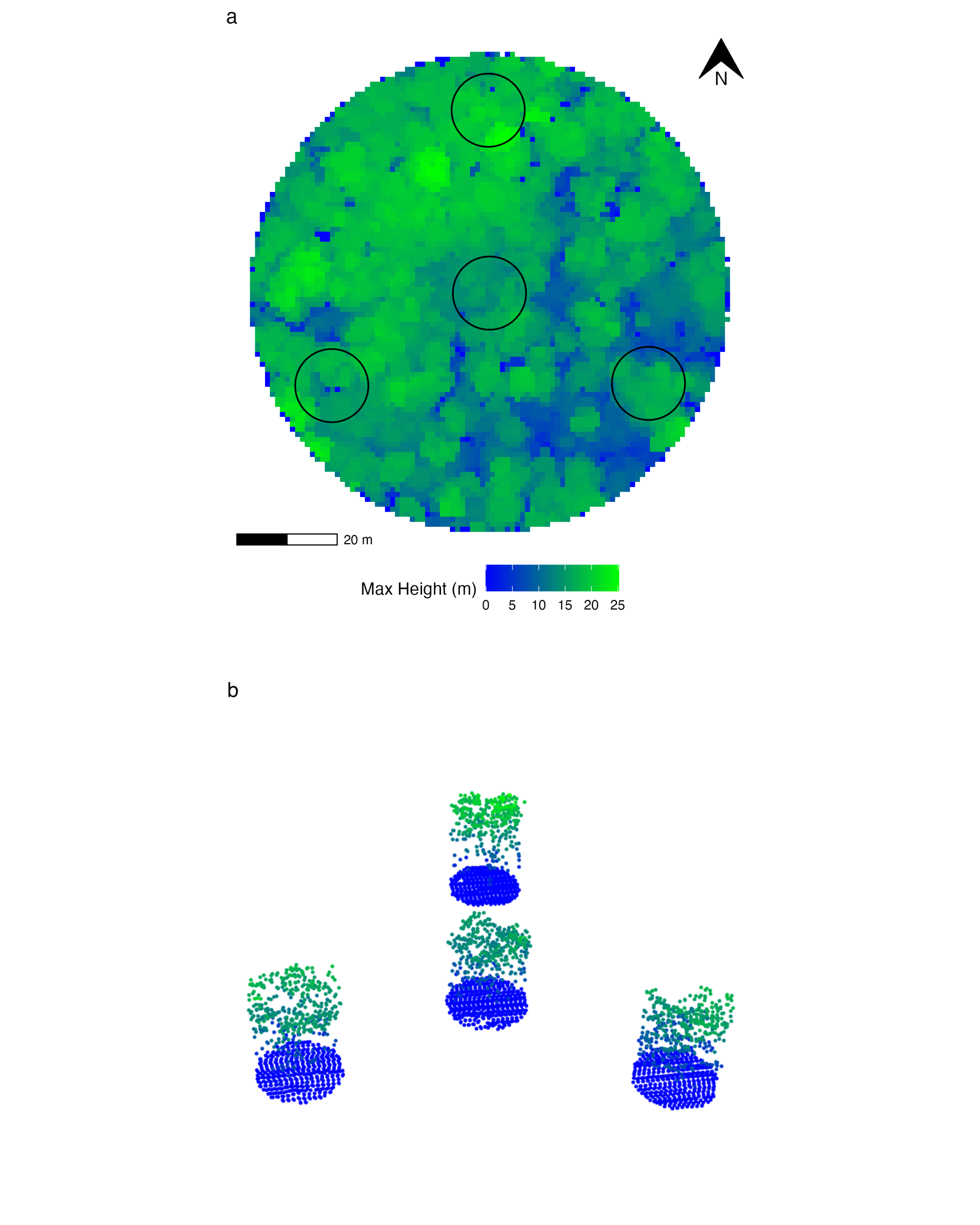} \caption{a) FIA plot layout overlaid on a 1 m LiDAR-derived max-height surface. b) Height-normalized LiDAR returns clipped to individual FIA subplots and pooled prior to predictor computation.}\label{fig:lasclip}
\end{figure}

FIA plots are composed of four identical circular subplots with radii of 7.32 m
(24 ft), with one subplot centered at the macroplot centroid and three
subplots located 36.6 m (120 ft) away at azimuths of 360\(^{\circ}\),
120\(^{\circ}\), and 240\(^{\circ}\) (Bechtold and Patterson 2005).
The plot locations were provided by the FIA program in the form of average
coordinates,
collected over multiple repeat visits,
representing the centroid of the center subplot,
which we then used to build a polygon dataset representing the entire plot
layout including all four subplots (Figure \ref{fig:lasclip}).
Averaged coordinates were necessary due to the lack of precision of
initial GPS coordinates for the macroplot centroids (Hoppus and Lister 2005).
Any reference to an FIA plot hereafter implies the aggregation of all four
component subplots.

\hypertarget{model-development-dataset}{%
\subsubsection{Model Development Dataset}\label{model-development-dataset}}

In selecting plots for model development
(model dataset hereafter; Table \ref{tab:lidarsummary})
we aimed to
maximize the number of reference plots (Fassnacht et al. 2014),
while minimizing the temporal lag between LiDAR acquisitions and inventories
and ensuring high quality co-registered LiDAR data
(White et al. 2013; CEOS 2021).
Temporal misalignment between plots and LiDAR has been shown to introduce
error in model predictions;
Gonçalves et al. (2017) found that 7-17\% of their prediction error could be attributed to
3-year lags between inventories and LiDAR acquisitions.

We only considered FIA plots when all subplots were marked
as measured and when forest conditions were uniform at the plot level
(all subplots completely forested or completely non-forested).
Only
324
plots were available when we required a strict temporal match between
inventories and LiDAR acquisitions
(Table \ref{tab:modeldataset}; Criterion 1).
To increase the number of reference plots,
we applied a growth adjustment to FIA plots with inventories
both before and after the LiDAR acquisition, or `bracketing inventories'.
For plots with bracketing inventories, we excluded those where AGB decreased
by \(\geq\) 5\% between measurements indicating a disturbance event on the plot.
For the remaining (relatively undisturbed) plots with bracketing inventories,
AGB at the time of the LiDAR collection was then estimated by linearly
interpolating between bracketing FIA estimates.
This procedure added
843
plots to our dataset (Table \ref{tab:modeldataset}; Criterion 2),
and is analogous to existing growth adjustment methods
(Gonçalves et al. 2017; Gobakken and Næsset 2008).

Several plot exclusion rules were implemented to filter duplicates and remove
problematic observations resulting from vegetation in non-forest conditions,
interfering structures, or other data anomalies which would otherwise
degrade the relationships between AGB and our predictor variables
(Table \ref{tab:modeldataset}; Criteria 3-5).
Point clouds were clipped to the constructed plot footprints and
were excluded with criteria 3-5 (Figure \ref{fig:lasclip}).

\hypertarget{map-assessment-dataset}{%
\subsubsection{Map Assessment Dataset}\label{map-assessment-dataset}}

For our map assessment
dataset (assessment dataset hereafter; Table \ref{tab:lidarsummary})
we were primarily concerned with maintaining
FIA's probability sample of inventory plots in order to leverage
unbiased estimators of map agreement metrics (Stehman and Foody 2019; Riemann et al. 2010).
As in the model dataset, only those plots with all four subplots marked as
measured were considered.
Excluding non-measured plots, however, does not invalidate FIA's probability
sample, as the FIA program assumes these plots to be randomly distributed across
the landscape (Bechtold and Patterson 2005).
To this end, we selected FIA plots on a coverage-by-coverage basis
(Figure \ref{fig:regions}),
choosing only plots that had been inventoried in +/- 2 years from the year of
the LiDAR acquisition,
but not inventoried in the same year as the LiDAR acquisition to maintain
independence from the model dataset
(Table \ref{tab:assessmentdataset}; Criterion 1).
Where plots were shared with the model dataset,
AGB values differed,
with AGB values derived from measurements in the map assessment dataset,
and AGB values derived from our growth-adjustment process in the model dataset,
therefore representing different points in time.
Furthermore, plots falling outside of our mapped area,
based on our landcover and area-of-applicability masks
(Section \ref{lcmapmethods}),
were excluded as they were considered outside our population of interest
(Table \ref{tab:assessmentdataset}; Criteria 2-3).

\begin{table}

\caption{\label{tab:modeldataset}Summary of plot selection criteria for the model dataset.}
\centering
\fontsize{10}{12}\selectfont
\begin{tabular}[t]{lr>{\raggedright\arraybackslash}p{14em}r}
\toprule
\multicolumn{1}{c}{} & \multicolumn{1}{c}{Criteria} & \multicolumn{1}{c}{Description} & \multicolumn{1}{c}{Num Plots}\\
\midrule
 & 1 & Plots inventoried in the same year as a LiDAR acquisition. & 324\\

\multirow[t]{-2}{*}{\raggedright\arraybackslash Include} & 2 & Plots growth-adjusted to temporally align with LiDAR acquisition. & 843\\
\addlinespace
\midrule
 & 3 & Plots where measured AGB = 0 Mg ha$^{-1}$ but maximum LiDAR return $>$ 1m. & 327\\

 & 4 & Plots where the convex hull of LiDAR returns clipped to a given subplot did not contain at least 90\% of the subplot area. & 13\\

\multirow[t]{-3}{*}{\raggedright\arraybackslash Exclude} & 5 & Plots duplicated in neighboring coverages. Temporal matches given preference, followed by inventory recency. & 26\\
\bottomrule
\end{tabular}
\end{table}

\begin{table}

\caption{\label{tab:assessmentdataset}Summary of plot selection criteria for the assessment dataset.}
\centering
\fontsize{10}{12}\selectfont
\begin{tabular}[t]{lr>{\raggedright\arraybackslash}p{14em}r}
\toprule
\multicolumn{1}{c}{} & \multicolumn{1}{c}{Criteria} & \multicolumn{1}{c}{Description} & \multicolumn{1}{c}{Num Plots}\\
\midrule
Include & 1 & Plots inventoried in +/- 2 years of LiDAR acquisition but not in the same year as the acquisition. & 1639\\
\addlinespace
\midrule
 & 2 & Plots with any portion of the plot footprint outside of our landcover mask. & 291\\

\multirow[t]{-2}{*}{\raggedright\arraybackslash Exclude} & 3 & Plots with any portion of the plot footprint outside of our area of applicability mask. & 131\\
\bottomrule
\end{tabular}
\end{table}

\hypertarget{lidar-and-auxiliary-data-pre-processing}{%
\subsection{LiDAR and Auxiliary Data Pre-Processing}\label{lidar-and-auxiliary-data-pre-processing}}

First, we height-normalized the raw LiDAR data and computed 40 predictors
(Supplementary Materials S1)
based on previous studies (Hawbaker et al. 2010; Huang et al. 2019; Pflugmacher et al. 2014).
Predictors at individual FIA plots were computed from LiDAR returns clipped to
only the measured subplot footprints and then pooled at the plot level
(Figure \ref{fig:lasclip}).
The corresponding predictors computed for map pixels were based on the set
of returns within each 30 m cell.
The lidR (Roussel and Auty 2020; Jean-Romain et al. 2020) package in R (R Core Team 2021) was used for
height-normalization and predictor generation.

A group of steady-state predictors was included to represent geospatial
variation in climate and topography (Kennedy et al. 2018).
Additionally, a 2019 tax parcel layer was incorporated as a set of
boolean indicator variables (Supplementary Materials S2).
Tax codes and categories provide cadastral information related to land-use and
management (Thompson et al. 2011). All auxiliary predictors were reprojected and
resampled to pixel geometries matching the 30 m LiDAR predictor surfaces.

\hypertarget{model-development}{%
\subsection{Model Development}\label{model-development}}

Three ML models were fit to a randomly selected 80\%
of the model dataset
(training set; n = 630),
leaving the 20\%
remaining plots as an independent testing set to assess the final model
performance on point clouds clipped to FIA plot footprints
(testing set; n = 171).
Random forest models (RF; ranger; Wright and Ziegler (2017)),
stochastic gradient boosting machines
(GBM; lightgbm; Ke et al. (2021), Ke et al. (2017)),
and support vector machines (SVM; kernlab; Karatzoglou et al. (2004))
were tuned using the 80\% training dataset and an iterative grid search approach,
starting by testing wide ranges of hyperparameters using five-fold cross
validation and then narrowing down to only the most performant combinations
over several iterations.
Models then used the most accurate sets of hyperparameters in all other
analyses.
For each of the \(n\) observation in the training dataset,
all three component models were fit, using their optimal hyperparameters,
with \(n-1\) observations.
Predictions for each component model were made for the nth (left out)
observation.
A linear regression model was used to estimate AGB as a function of these
leave-one-out predictions,
combining the three ML approaches in a stacked ensemble
to reduce the generalization error of our component models (Wolpert 1992).

\hypertarget{modelperf}{%
\subsection{Model Performance}\label{modelperf}}

Model performance was assessed against the 20\% testing partition of the model
dataset (Table \ref{tab:lidarsummary})
based on metrics including
root-mean-squared error in Mg ha\textsuperscript{-1}
(RMSE, equation \eqref{eq:rmse}),
percent RMSE (\% RMSE, equation \eqref{eq:prmse}),
mean absolute error in Mg ha\textsuperscript{-1} (MAE, equation \eqref{eq:mae}),
percent MAE (\% MAE, equation \eqref{eq:pmae}),
mean error in Mg ha\textsuperscript{-1} (ME, equation \eqref{eq:me}),
and the coefficient of determination (\(R^2\), equation \eqref{eq:r2}) as follows:

\begin{equation}
\operatorname{RMSE} = \sqrt{(\frac{1}{n})\sum_{i=1}^{n}(y_{i} - \hat{y_{i}})^{2}} \label{eq:rmse}
\end{equation}

\begin{equation}
\operatorname{\%\ RMSE} = 100 \cdot \frac{\operatorname{RMSE}}{\bar{y}} \label{eq:prmse}
\end{equation}

\begin{equation}
\operatorname{MAE} =  (\frac{1}{n})\sum_{i=1}^{n}(\lvert y_{i} - \hat{y_{i}} \rvert) \label{eq:mae}
\end{equation}

\begin{equation}
\operatorname{\%\ MAE} = 100 \cdot \frac{\operatorname{MAE}}{\bar{y}} \label{eq:pmae}
\end{equation}

\begin{equation}
\operatorname{ME} = (\frac{1}{n})\sum_{i=1}^{n}(y_{i} - \hat{y_{i}}) \label{eq:me}
\end{equation}

\begin{equation}
\operatorname{R^2} = 1 - \frac{\sum_{i=1}^{n}\left(y_{i}-\hat{y}_{i}\right)^2}{\sum_{i=1}^{n}\left(y_i - \bar{y}\right)^2} \label{eq:r2}
\end{equation}

\noindent where \(n\) is the number of FIA plots included in the data set,
\(\hat{y_i}\) is the predicted value of AGB, \(y_{i}\) the AGB value measured at
the corresponding location,
and \(\bar{y}\) the mean AGB value from FIA field measurements.

Standard errors for R\textsuperscript{2} and RMSE were computed as follows:
\begin{equation}
\operatorname{SE_{boot}} = \sqrt{\frac{Var_{boot}}{n}} \label{eq:bootse}
\end{equation}

\noindent where \(n\) is the number of FIA plots included in the dataset, and
\(Var_{boot}\) is computed as the variance of R\textsuperscript{2}
and RMSE estimates for 1000 iterations of bootstrap resampling.
Standard errors for MAE and ME were computed as follows:

\begin{equation}
\operatorname{SE} = \sqrt{\frac{\sum_{i=1}^{n}\left(e_i - \bar{e}\right)^2}{n - 1}} \label{eq:se}
\end{equation}

\noindent where \(e_{i}\) is the error at an FIA plot,
and \(\bar{e}\) is the~mean error from all FIA plots included.
In the case of MAE, \(e_{i} = |y_{i} - \hat{y_{i}}|\), the absolute value of the
error at a given FIA plot.

\hypertarget{lcmapmethods}{%
\subsection{AGB Mapping and Post-processing}\label{lcmapmethods}}

The linear model ensemble was used to make predictions for all
30 m pixels within the GPO-LiDAR area.
Predictions on newer coverages superseded those based on older coverages in
areas where neighboring LiDAR coverages overlapped.
All AGB prediction surfaces were projected to match Landsat 30 m pixel
geometries to avoid mixed pixel effects in subsequent raster overlay analyses
(Wulder et al. 2022).

With recognition that our predictions are best suited to areas populated by
woody biomass, we tabulated our predictions across landcover types determined
from the United States Geological Survey's
Land Change Monitoring, Assessment, and Projection (LCMAP) primary
classification products (Brown et al. 2020; Zhu and Woodcock 2014), which has a reported
overall accuracy of 77.4\% in the eastern United States for the years 1985-2018 (Pengra et al. 2020).
LCMAP's annual resolution (1985-2019) allowed for temporal alignment with the
patchwork of LiDAR-AGB surfaces, and shared identical pixel geometries.
We masked our AGB prediction surfaces to remove
Developed, Water, and Barren pixels, and then tabulated AGB by the four
remaining vegetated LCMAP classes of Tree Cover, Grass/Shrub, Cropland,
and Wetland.

Lastly, we computed an area of applicability (AOA) surface which
identifies pixels containing predictor data that are beyond a pre-specified
predictor-space distance from the training data
(Meyer and Pebesma (2021); Supplementary Materials S4).
We used this surface as a final mask for our AGB prediction surfaces,
restricting our maps to only areas that were well-represented in the training
data.

\hypertarget{methodsriemann}{%
\subsection{Map Agreement Analysis}\label{methodsriemann}}

We assessed the agreement between our AGB maps and FIA
reference data following approaches prescribed by Riemann et al. (2010) and
Menlove and Healey (2020).
The former evaluated agreement across a range of scales and accounts
for the mismatch in spatial support between map
aggregate estimates (many pixels) and FIA aggregate estimates (few plots) by
only extracting pixels coincident with FIA plots.
The latter compared FIA-derived AGB estimates ---
which have been adjusted for forest cover within, and area-extrapolated to,
hexagon map units ---
to zonal averages of our mapped AGB.

Following Riemann et al. (2010) we compared our AGB prediction surfaces
to the assessment dataset (Section \ref{field-data}).
Comparisons were made at both the plot-to-pixel
scale and within variably-sized hexagons with distances between centroids
ranging from 10 km
(8660 ha)
to 100 km
(866,025 ha).
As an extension of the Riemann et al. (2010) methodology we assessed prediction error
(RMSE, MAE, ME) with choropleth maps that summarized the mapped
residuals and FIA reference data distributions within hexagon units with
centroids spaced 50 km apart.
We also grouped plot to pixel results by the majority LCMAP classification
at each plot, to demonstrate the level of agreement across vegetated
landcover classes.
To evaluate the presence of spatial autocorrelation among mapped residuals,
which could indicate the presence of regional or coverage specific error in
our model's predictions,
Global Moran's I statistics were computed for search radii ranging from
1 to 50 km using both the model dataset and the assessment dataset
separately (Moran 1950).

Following the Menlove and Healey (2020) approach we compared the average of our masked
predictions, weighted by the proportion of each pixel intersecting a given
hexagon, to a set of FIA-derived estimates for 64,000 ha hexagons representing
FIA's finest acceptable scale for the most recent inventory cycle in NYS
(2013-2019) (Menlove and Healey 2020).
As recommended by Menlove and Healey (2020), we accounted for differences in forest
definitions between the FIA estimates and our mapped estimates by dividing FIA
estimates by the total area of vegetated
(based on LCMAP Tree cover, Wetland, Cropland, Grass/Shrub)
pixels within each hexagon.
Lastly, we limited this comparison to only those hexagons that contained \textgreater{}
10\% mapped area, using the summed area of all pixels inside of our LCMAP and
AOA mask
(Section \ref{lcmapmethods})
to define mapped area within each hexagon.

The exactextractr (Daniel Baston 2021), spdep (Bivand, Pebesma, and Gomez-Rubio 2013), sf (Pebesma 2018),
raster (Hijmans 2021), and terra (Hijmans 2022) packages in R (R Core Team 2021) were used to conduct
all analyses described here.
Further description of this assessment is included in Supplementary Materials
S5.

\hypertarget{results}{%
\section{Results}\label{results}}

\hypertarget{growth-adjusted-field-plots}{%
\subsection{Growth-Adjusted Field Plots}\label{growth-adjusted-field-plots}}

Of the
801
FIA plots in our model dataset,
562
were growth-adjusted, representing about
70\%
of the full dataset.
The average annual AGB increment applied to all growth-adjusted plots
was
1.84
Mg ha\textsuperscript{-1} year\textsuperscript{-1} over an average of
3.96
years.
The largest annual AGB increment applied was
11.17
Mg ha\textsuperscript{-1} year\textsuperscript{-1} over
3
years.
The longest adjustments (applied to
5
plots) occurred over
10
years at an average annual increment of
2.63
Mg ha\textsuperscript{-1} year\textsuperscript{-1}.

\hypertarget{model-performance}{%
\subsection{Model Performance}\label{model-performance}}

\begin{figure}
\includegraphics[width=0.75\linewidth]{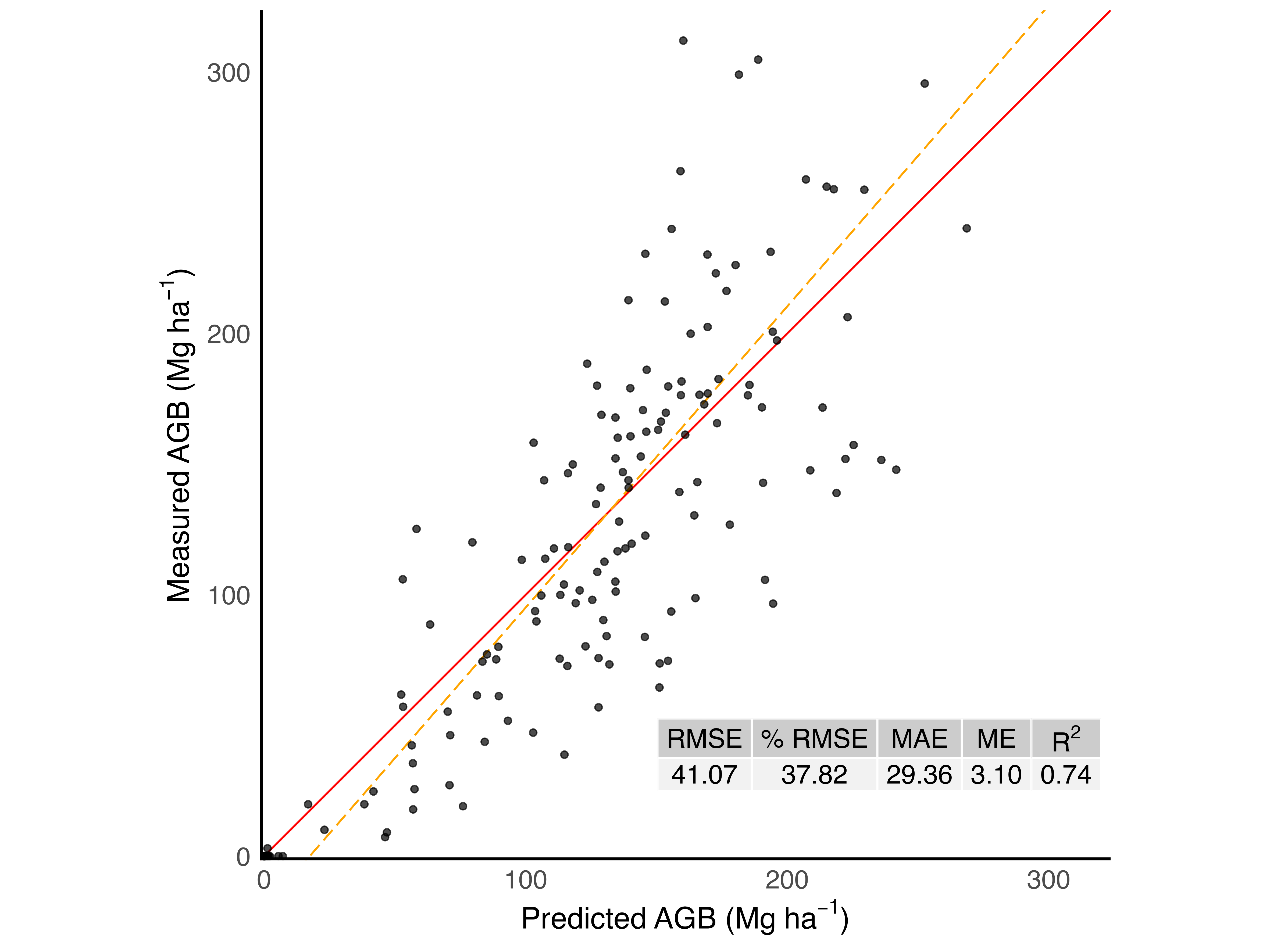} \caption{Measured vs predicted AGB scatter plot for the testing portion of the model dataset. AGB values in Mg ha$^{-1}$. GMFR trend line shown with dashed (orange) line, and 1:1 line shown with solid (red) line}\label{fig:modelperfscatter}
\end{figure}

The final model performance against the 20\% test partition of the model dataset
showed generally accurate predictions, but with greater accuracy towards the
mean of the reference distribution (Figure \ref{fig:modelperfscatter}).
This resulted in slight overpredictions at smaller AGB plots
and underpredictions at larger AGB plots (Figure \ref{fig:modelperfscatter}).
A more detailed analysis of performance for the ensemble model and the three
component models can be found in Supplementary Materials
S3.

\hypertarget{agb-by-landcover-class}{%
\subsection{AGB by Landcover Class}\label{agb-by-landcover-class}}

\begin{figure}
\includegraphics[width=1\linewidth]{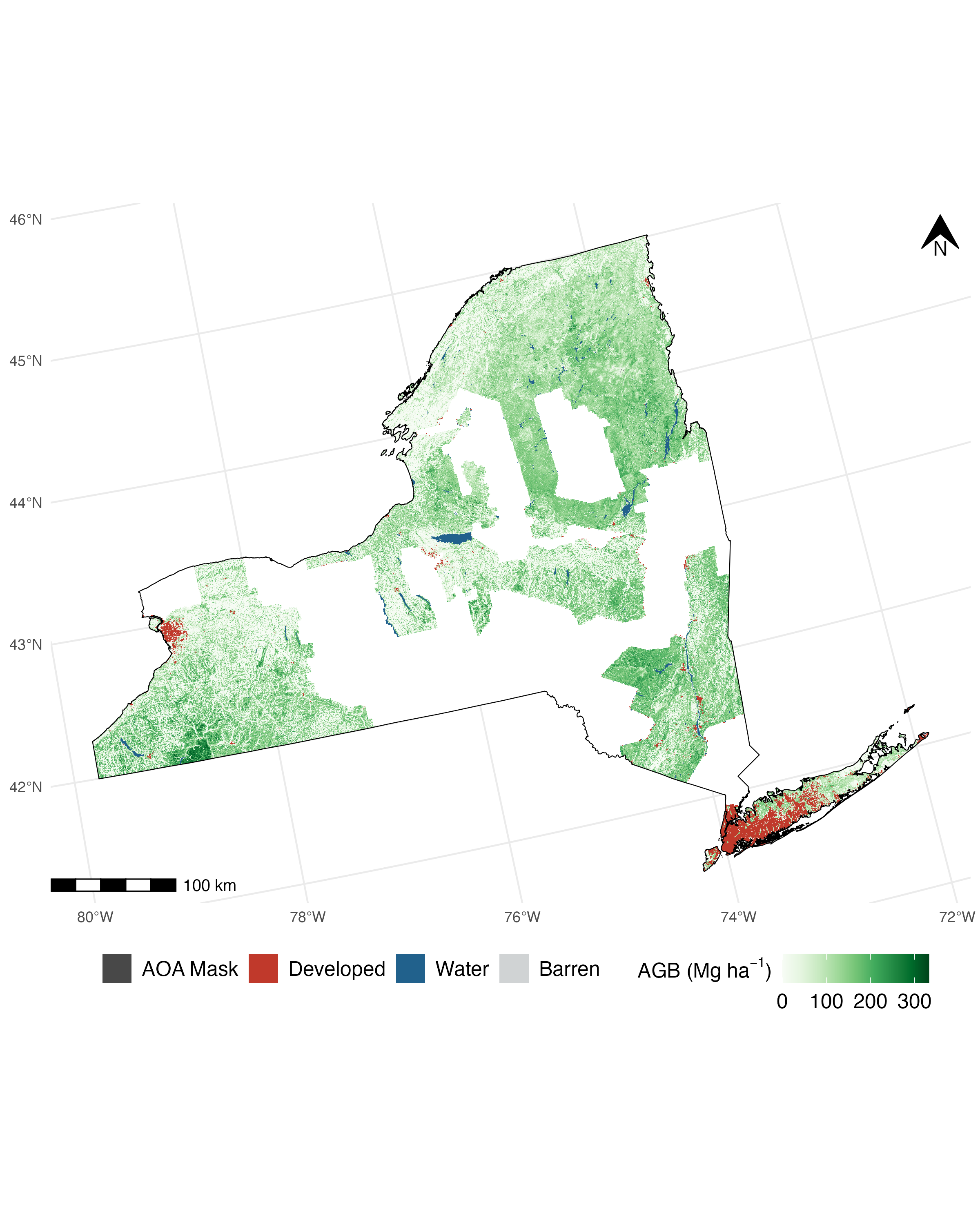} \caption{AGB prediction surfaces reflecting a temporal patchwork of conditions across the 17 component LiDAR coverages in the GPO-LiDAR area with New York State (USA).}\label{fig:maps}
\end{figure}

Approximately
62\%
of all pixels, accounting for
87\%
of the total mapped AGB (Figure \ref{fig:maps}),
were contained within areas identified as Tree cover by LCMAP
(Table \ref{tab:binning}).
All other LCMAP classes contained comparatively small, but non-zero estimates
of AGB, with Wetlands and Grass/Shrub containing larger average predictions
within small portions of the mapped area,
and Cropland containing smaller average predictions across a larger proportion
of the mapped area (Table \ref{tab:binning}).

\begin{table}

\caption{\label{tab:binning}Summary of FIA reference plots in the model dataset and mapped predictions by LCMAP landcover classes. n = number of plots. Mean AGB values in Mg ha$^{-1}$. Total AGB values in millions of metric tons. Area in hectares. AOA expressed as percent of LCMAP classified pixels considered inside of the area of applicability.}
\centering
\begin{tabular}[t]{lrrrrrrrr}
\toprule
\multicolumn{1}{c}{ } & \multicolumn{2}{c}{Reference Plots} & \multicolumn{6}{c}{Mapped} \\
\cmidrule(l{3pt}r{3pt}){2-3} \cmidrule(l{3pt}r{3pt}){4-9}
\multicolumn{1}{c}{LCMAP} & \multicolumn{1}{c}{n} & \multicolumn{1}{c}{Mean AGB} & \multicolumn{1}{c}{Area} & \multicolumn{1}{c}{\% Area} & \multicolumn{1}{c}{Mean AGB} & \multicolumn{1}{c}{Total AGB} & \multicolumn{1}{c}{\% AGB} & \multicolumn{1}{c}{\% AOA}\\
\midrule
Tree cover & 599 & 137.97 & 4,251,812 & 62.42 & 132.66 & 564.06 & 87.38 & 98.13\\
\addlinespace
Cropland & 132 & 1.98 & 1,748,905 & 25.68 & 14.11 & 24.67 & 3.82 & 98.25\\
\addlinespace
Wetland & 36 & 115.13 & 613,578 & 9.01 & 77.22 & 47.38 & 7.34 & 97.64\\
\addlinespace
Grass/Shrub & 12 & 30.68 & 196,852 & 2.89 & 47.86 & 9.42 & 1.46 & 98.17\\
\bottomrule
\end{tabular}
\end{table}

\hypertarget{area-of-applicability}{%
\subsection{Area of Applicability}\label{area-of-applicability}}

All mapped (vegetated) LCMAP classes contained AOA \(\geq\)
97.64\%
and the AOA breakdown across component LiDAR coverages (after LCMAP masking)
was uniform
(\(\geq\)
90.4\%)
with NYC and Long Island the only two coverages with \(\leq\) 95\% AOA
(Tables \ref{tab:lidarsummary} and \ref{tab:binning}).
In total,
98.12\%
of the GPO-LiDAR area was considered inside the AOA after initial LCMAP
masking.

\hypertarget{map-agreement-analysis}{%
\subsection{Map Agreement Analysis}\label{map-agreement-analysis}}

\begin{figure}
\includegraphics[width=1\linewidth]{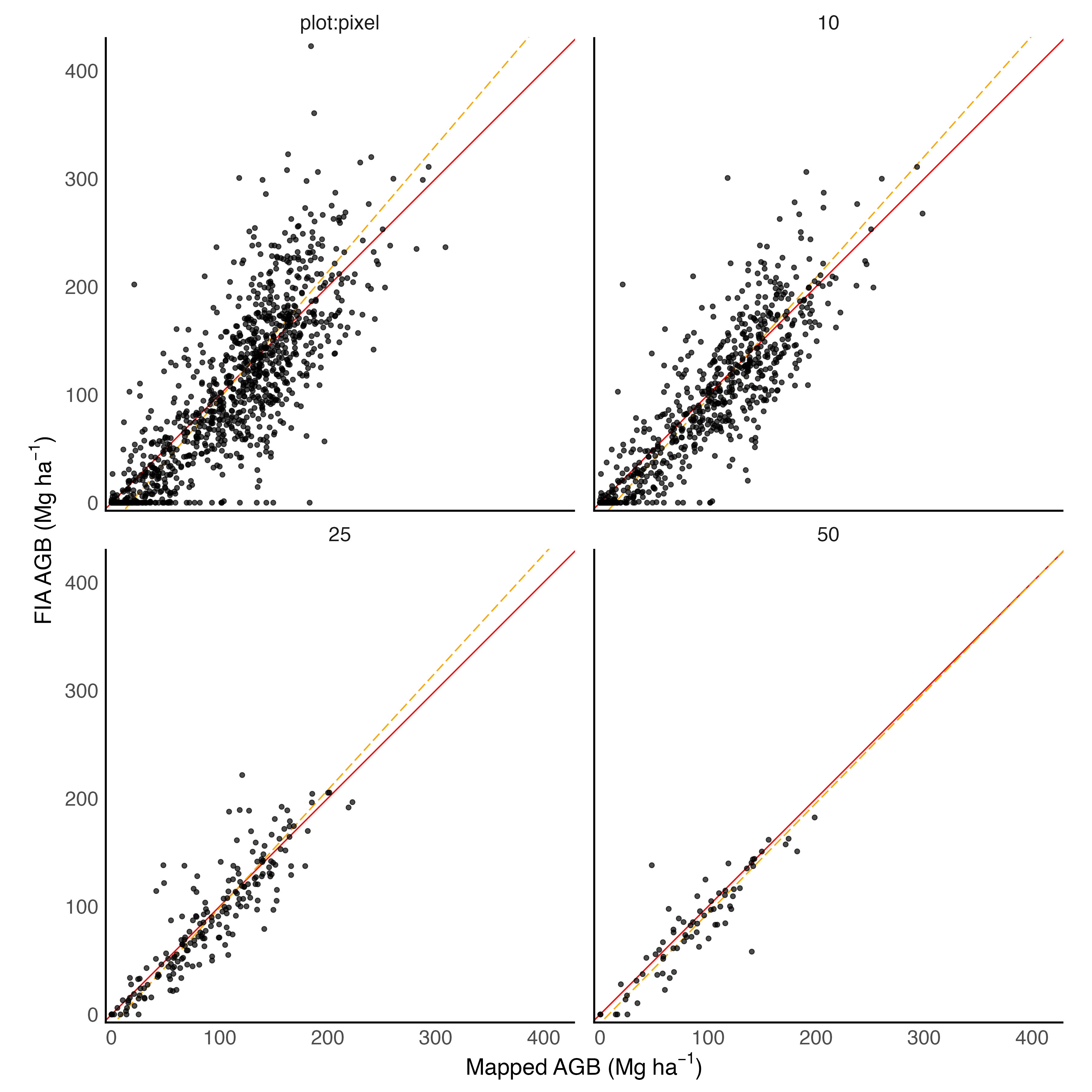} \caption{Comparing mapped AGB to FIA estimated AGB across selected scales represented by distances between hexagon centroids (plot:pixel, 10 km, 25 km, and 50 km). AGB values in Mg ha$^{-1}$. GMFR trend line shown with dashed (orange) line, and 1:1 line shown with solid (red) line.}\label{fig:riemann11}
\end{figure}

\begin{figure}
\includegraphics[width=1\linewidth]{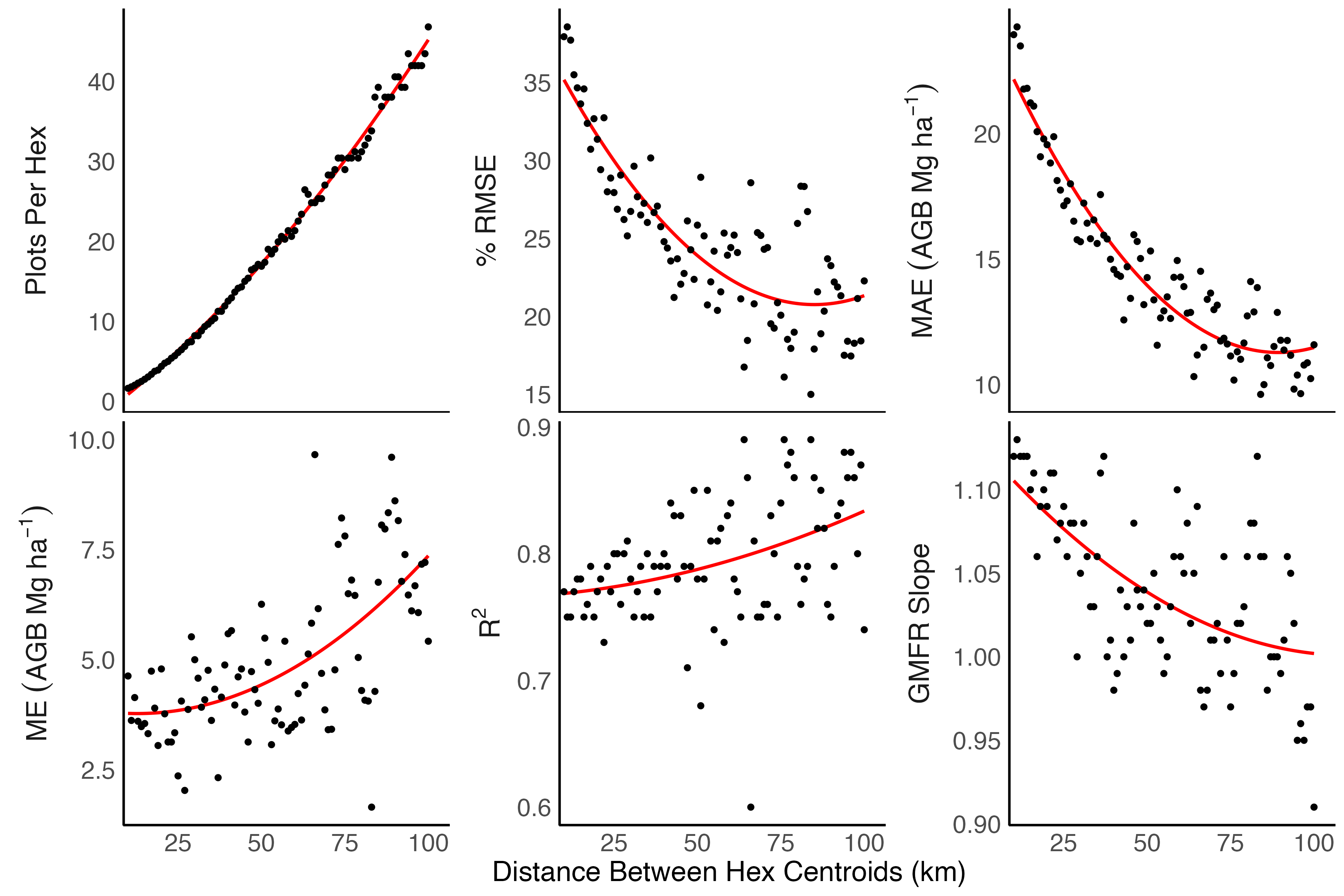} \caption{Summary assessment metrics (as defined in Section 2.5) comparing mapped predictions to FIA estimates as a function of aggregation unit size (described by distances between hexagon centroids). Red trend lines produced using quadratic regression.}\label{fig:riemanncont}
\end{figure}

We observed improved agreement between the mapped estimates and the FIA
estimates as the aggregation unit size increased, with \% RMSE decreasing from
45 to
15\%
and MAE decreasing from
28 to
10
Mg ha\textsuperscript{-1} (Table \ref{tab:acctable}, Figures \ref{fig:riemann11} and
\ref{fig:riemanncont}).
The scatter plot results (Figure \ref{fig:riemann11}) for the plot to pixel
and 10 km scales show a pattern of large overpredictions
where FIA reference values are 0 Mg ha\textsuperscript{-1} AGB --- likely an indication of the
structural zeroes introduced by FIA's strict definition of forest.
Notably, ME increased as a function of aggregation unit size, which appears
to be a reflection of outlier hexagons with very small samples rather than
any additive effects from aggregation
(Supplementary materials S5).
We again observed less accurate predictions at the extremes of the distributions
across most scales, though our extreme predictions (small and large) became more
accurate as a function of aggregation unit size, as indicated by the
geometric mean functional relationship (GMFR; Riemann et al. (2010)) slope approaching
1 at the largest scales
(Figures \ref{fig:riemann11} and \ref{fig:riemanncont}).

\begin{figure}

{\centering \includegraphics[width=1\linewidth]{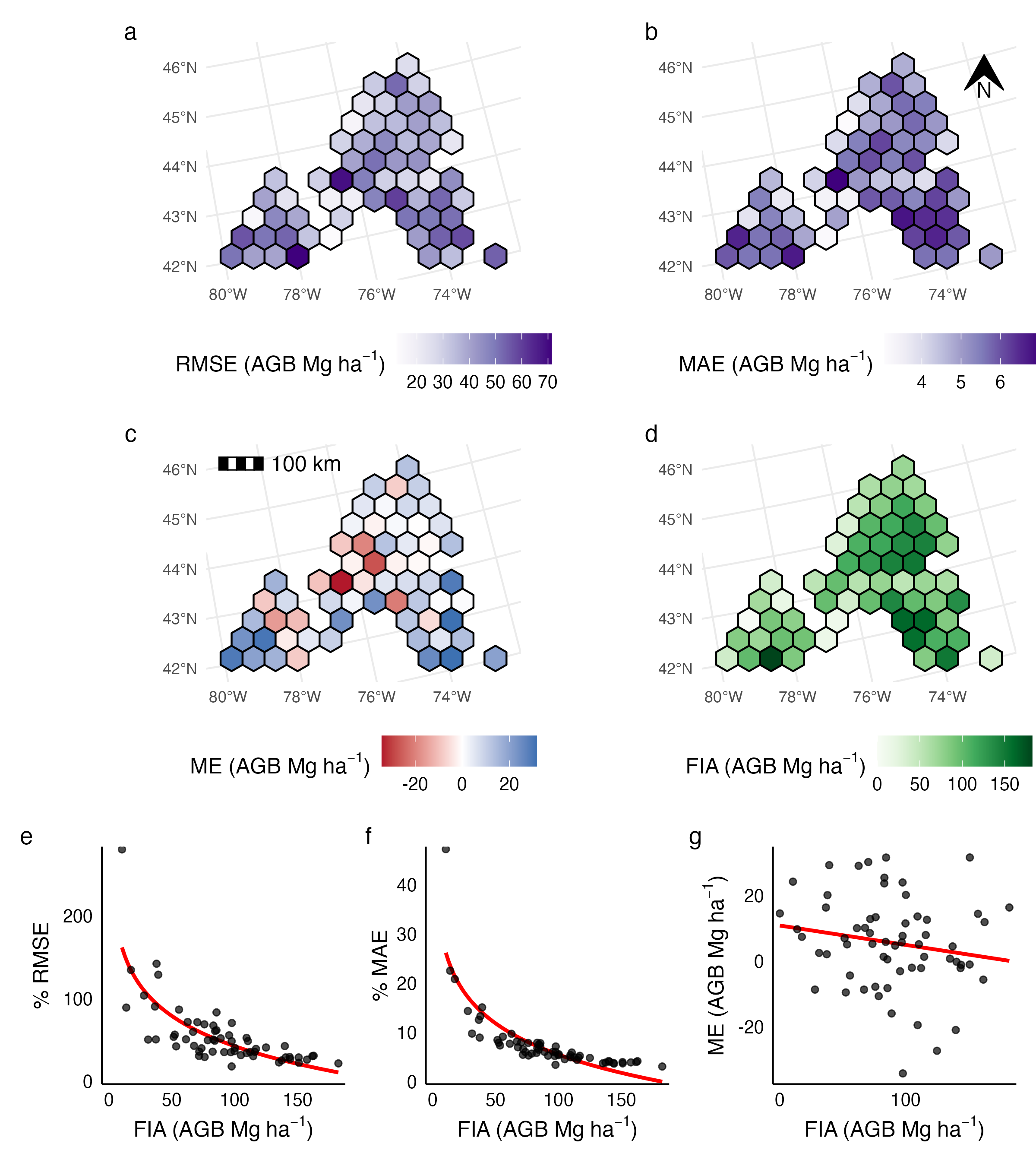} 

}

\caption{Plot-to-pixel residuals summarized at units spaced 50 km apart. Hexagons with only one reference plot were removed. a) RMSE Mg ha$^{-1}$ b) MAE Mg ha$^{-1}$ c) ME Mg ha$^{-1}$ d) Mean FIA AGB value Mg ha$^{-1}$ e) Hex-level \% RMSE as a function of mean reference value f) Hex-level \% MAE as a function of mean reference value. g) Hex-level ME as a function of mean reference value within each hexagon. 1 observation excluded from e and f where 0 FIA AGB makes \% RMSE impossible to compute. Trend lines in e and f produced using logarithmic regression. Trend line in g produced using least-squares regression. RMSE, MAE, and ME as defined in Section 2.5.}\label{fig:hexpanel}
\end{figure}

The model's mapped residuals grouped to 50 km spaced hexagons did not reveal
any observable spatial patterns of prediction error (RMSE, MAE, ME)
(Figure \ref{fig:hexpanel}).
The ME map (Figure \ref{fig:hexpanel}c) was dominated by near-zero positive
prediction error but indicated the tendency for hexagons with negative
prediction ME to have larger mean FIA AGB values,
while hexagons with positive ME were more likely to have smaller
mean FIA AGB values (Figure \ref{fig:hexpanel}g).
The RMSE map (Figure \ref{fig:hexpanel}a) and MAE map
(Figure \ref{fig:hexpanel}b) largely mirrored one another.

\begin{figure}

{\centering \includegraphics[width=1\linewidth]{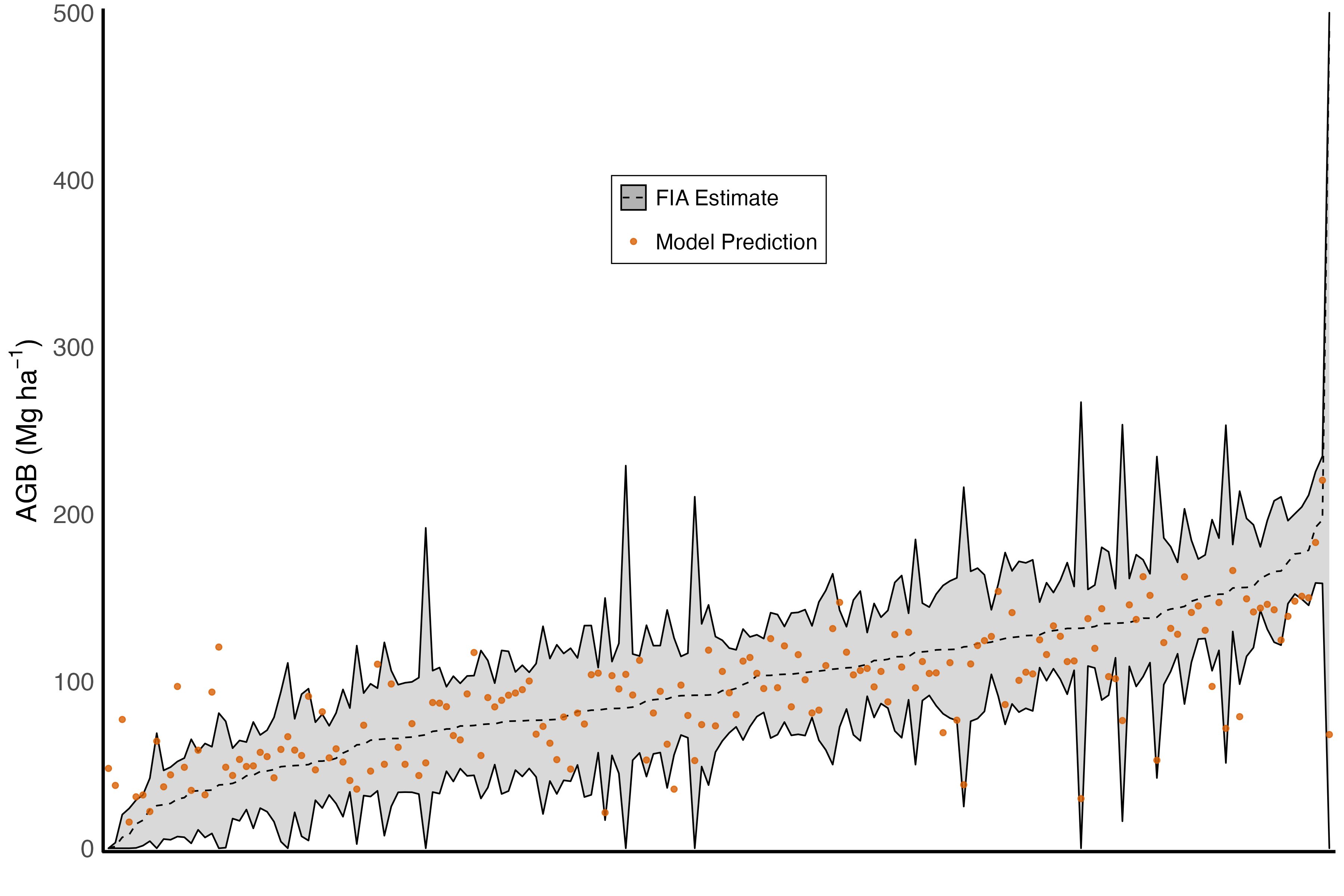} 

}

\caption{Comparison of mapped predicted AGB to Menlove and Healey (2020) estimates (dashed line) and associated 95\% confidence interval (gray shaded region) within 64,000-ha aggregation hexagons. FIA estimates of AGB are scaled by the proportion of forest cover indicated by LCMAP 2016 Tree cover, Wetland, Croplands, and Grass/Shrub classified pixels. Hexagons with mapped areas  $\leq$ 10\% of their total area were excluded from this analysis.  Observations are sorted by increasing FIA estimates along the x-axis.}\label{fig:standardfia}
\end{figure}

Comparison of our AGB maps with FIA's design based estimates of AGB density
(Menlove and Healey 2020) indicated overall strong agreement, with
89\%
of our estimates falling within the FIA estimate 95\% confidence intervals (CI)
(Figure \ref{fig:standardfia}).
The majority of our estimates falling outside the CI were at the lower range of
the AGB distribution,
reinforcing the observed pattern of overprediction in comparison to
FIA-derived estimates at the low end of the reference distribution.

Map agreement across LCMAP vegetated classes was highly variable,
with the smallest \% RMSE for plots classified as Tree Cover by LCMAP
(Table \ref{tab:lcpriacc}).
Absolute RMSE for Grass/Shrub and Cropland plots was small (\textless{} 24 Mg ha\textsuperscript{-1}),
but their relative values were quite large (\% RMSE \textgreater{} 100\%), reflecting the
small FIA AGB averages within these classes.
Mean errors were positive and large for Grass/Shrub and Croplands
(14.64 Mg ha\textsuperscript{-1} and
6.67 Mg ha\textsuperscript{-1} respectively; Table \ref{tab:lcpriacc}),
suggesting that errors contained in these two landcover classes had a
relatively large contribution to the overall patterns of positive ME and
overprediction on the low end of the FIA AGB distribution across the entire
GPO-LiDAR area.

The global Moran's I analysis with the assessment dataset found evidence of
very weak spatial autocorrelation (\(\leq\) 0.10) in mapped residuals for
all search radii (Supplementary Materials S6).
Notably, when the analysis was conducted with the model dataset,
similarly very weak spatial autocorrelation (\(\leq\) 0.08) was only evident for
search radii of 9 km to 14 km, but was not evident at any other scales
(Supplementary Materials S6).

\begin{table}

\caption{\label{tab:acctable}Map agreement results for select scales. Distance = distance between hexagon centroids in km; PPH = plots per hexagon; n = number of comparison units (plots or hexagons); RMSE, MAE, ME in Mg ha$^{-1}$. All accuracy metrics as defined in Section 2.5. Standard errors in parentheses. }
\centering
\begin{tabular}[t]{rrrrrrrr}
\toprule
\multicolumn{1}{c}{Distance} & \multicolumn{1}{c}{n} & \multicolumn{1}{c}{PPH} & \multicolumn{1}{c}{\% RMSE} & \multicolumn{1}{c}{RMSE} & \multicolumn{1}{c}{MAE} & \multicolumn{1}{c}{ME} & \multicolumn{1}{c}{R$^2$}\\
\midrule
 & 1217 &  & 44.87 & 40.93 (0.04) & 28.12 (0.85) & 4.40 (1.17) & 0.73 (0.01)\\
\addlinespace
10 & 739 & 1.65 & 37.95 & 34.06 (0.05) & 23.95 (0.89) & 4.63 (1.24) & 0.77 (0.01)\\
\addlinespace
25 & 199 & 6.12 & 27.95 & 24.68 (0.13) & 17.13 (1.26) & 2.36 (1.75) & 0.80 (0.01)\\
\addlinespace
50 & 72 & 16.9 & 25.86 & 21.17 (0.41) & 14.26 (1.86) & 6.26 (2.40) & 0.78 (0.01)\\
\addlinespace
100 & 26 & 46.81 & 22.28 & 17.39 (0.71) & 11.58 (2.59) & 5.42 (3.30) & 0.74 (0.03)\\
\bottomrule
\end{tabular}
\end{table}

\begin{table}

\caption{\label{tab:lcpriacc}Map agreement at the plot to pixel scale, grouped by LCMAP classification. n = number of plots; RMSE, MAE, ME in Mg ha$^{-1}$. All accuracy metrics as defined in Section 2.5. Standard errors in parentheses, with a minimum of 0.01 for display.}
\centering
\begin{tabular}[t]{lrrrrrr}
\toprule
\multicolumn{1}{c}{LCMAP} & \multicolumn{1}{c}{n} & \multicolumn{1}{c}{\% RMSE} & \multicolumn{1}{c}{RMSE} & \multicolumn{1}{c}{MAE} & \multicolumn{1}{c}{ME} & \multicolumn{1}{c}{R$^2$}\\
\midrule
Tree Cover & 797 & 36.35 & 45.85 (2.56) & 34.42 (1.07) & 3.83 (1.62) & 0.55 (0.01)\\
Cropland & 303 & 229.82 & 18.90 (2.40) & 10.32 (0.91) & 6.67 (1.02) & 0.40 (0.03)\\
Wetland & 91 & 62.48 & 51.36 (30.77) & 35.56 (3.91) & -1.13 (5.41) & 0.55 (0.01)\\
Grass/Shrub & 26 & 119.65 & 23.71 (17.01) & 16.03 (3.49) & 14.64 (3.73) & 0.50 (1.24)\\
\bottomrule
\end{tabular}
\end{table}

\hypertarget{discussion}{%
\section{Discussion}\label{discussion}}

In this study we attempted to use a patchwork of 17 discrete LiDAR coverages for
broad-scale, fine-resolution forest aboveground biomass (AGB) mapping across
New York State (NYS), USA.
Faced with a limited sample of temporally aligned field inventory data,
we leveraged repeated inventories to boost the model training sample,
and used a machine learning ensemble model to produce accurate predictions.
We addressed concerns of sensor and mission discrepancies among
component LiDAR coverages by investigating spatial patterns of prediction error,
and by using an AOA mask to both show predictor uniformity across coverages,
as well as to mask predictions based on anomalous data.
Our results demonstrated that our maps accurately characterized the spatial
patterns of AGB across the state,
and showed that our predictions have strong agreement with FIA estimates
across a range of aggregation scales.

\hypertarget{growth-adjusted-field-plots-1}{%
\subsection{Growth-Adjusted Field Plots}\label{growth-adjusted-field-plots-1}}

Our approach to solve the common lack of temporally coincident
LiDAR and field data was parsimonious and effectively tripled the sample size
while achieving accurate modeling results.
We did this by leveraging the existing inventory data, without additional field
campaigns, remotely sensed data, or growth and yield models.
However, a potential limitation to this approach is the requirement of
regular historical inventories so that bracketing inventory years can be
identified for LiDAR acquisitions.
Additionally, the maximum temporal distance between growth-adjusted AGB
values and the nearest measured AGB values in our dataset was
10
years;
growth-adjustment following our approach may not be reliable for longer
temporal lags.

\hypertarget{pooled-modeling}{%
\subsection{Pooled Modeling}\label{pooled-modeling}}

Despite our efforts to boost the amount of training information available to
models via growth adjustment, we were left with a non-uniform
spatial arrangement of FIA plots across the GPO-LiDAR area.
Several coverages contained fewer than 20 plots in the model dataset
(Table \ref{tab:lidarsummary}), requiring a pooled modeling approach.
Pooling information from all component LiDAR coverages allowed our model to
borrow information from coverages with more training data to build relationships
between our predictors and AGB that were then applied to coverages with less
training data.

We relied on the computed AOA surface to enforce predictor space similarity with
the training data, thus ensuring that our model did not predict into unknown
predictor space,
even in coverages with limited FIA plots.
Moreover, the AOA surface provided evidence of predictor-space uniformity
across all 17 component LiDAR coverages,
indicating that each of the component coverages was well represented in the
model training dataset (Meyer and Pebesma 2021).
It stands to reason, however, that the NYC coverage contained the lowest
proportion of AOA, since there were only two model plots available in this
coverage (Table \ref{tab:lidarsummary}).
Generally, pixels falling outside the AOA surface appeared to be the result of
problems with LiDAR collection or data processing abnormalities,
with some visible outliers that could not be attributed to any known ecological
phenomena.
We found the AOA mapping especially valuable in utilizing publicly available
LiDAR coverages off the shelf with limited knowledge of, or responsibility for,
their provenance.

\hypertarget{forest-definition-disparities}{%
\subsection{Forest Definition Disparities}\label{forest-definition-disparities}}

A forest can be defined in many ways depending on goals, perspectives,
and operational concerns,
and aligning the various definitions to make comparisons or derive relationships
is neither a trivial nor a unique challenge
(Chazdon et al. 2016; Riemann et al. 2010; Huang et al. 2019).
FIA's strict forest definition that is in part based on field observations of
land-use,
which is traditionally difficult to classify with remotely sensed data
(Fritz et al. 2017),
was difficult to harmonize with LiDAR.
Since FIA does not provide a forest/nonforest map,
and FIA's definition of forest can exclude significant AGB stocks in areas
containing tree cover (Johnson et al. 2015, 2014),
we relied on LCMAP's classifications of vegetated cover types to mask our
prediction surfaces.
Our AGB maps thus reflect a more inclusive definition of forest than FIA,
incorporating AGB stocks across a broader range of conditions and land-uses.
Despite these definitional differences, there was overlap between FIA forest
and our LCMAP-derived definition,
as evidenced by the non-zero FIA AGB averages for model dataset plots grouped
within each of the four vegetated LCMAP classes (Table \ref{tab:binning}).
Further, we were able to separate true zeros from structural zeros in FIA
nonforest conditions using a 1 m LiDAR max-height threshold
(Table \ref{tab:modeldataset}), giving our model information to make
predictions in areas with little to no canopy cover.

Although most of our mapped AGB was contained within pixels classified as tree
cover,
a significant minority was contained within pixels classified as Cropland and
Grass/Shrub (Table \ref{tab:binning}).
As the total area of agricultural lands has been declining in NYS (USDA National Agricultural Statistics Service 2019),
we expect that many of these predictions represented early successional
forests,
or patches of `young and stunted trees' in open fields (Yang et al. 2018),
both transitional states which are challenges for LCMAP's classification
algorithm (Mahoney, Johnson, and Beier 2022; Brown et al. 2020).
In summary, the landscape-level context in NYS,
and LCMAP's algorithmic challenges provided support to generate predictions
for all LCMAP vegetated classes.

\hypertarget{model-performance-and-map-agreement-assessment}{%
\subsection{Model Performance and Map Agreement Assessment}\label{model-performance-and-map-agreement-assessment}}

Our prediction accuracy against the 20\% testing partition of our model dataset
was favorably comparable to previous LiDAR-AGB mapping studies
(Huang et al. 2019; Nilsson et al. 2017; Ayrey et al. 2021; Hauglin et al. 2021).
Using a set of FIA-developed methods (Riemann et al. 2010; Menlove and Healey 2020) we further
demonstrated an overall strong agreement between our map-based estimates and
FIA-derived estimates.

It is unsurprising that assessment metrics generally improved as the scales of
aggregation increased, given that the plot-to-pixel scale can be considered the
most rigorous, with the largest variance in both the pixel and plot AGB
distributions, as well as the most potential for spatial misalignment between
30 m pixel predictions and FIA plot measurements to influence agreement metrics
(McRoberts et al. 2018).
When mapped residuals were summarized within units spaced 50 km apart,
larger magnitudes of prediction error (RMSE, ME) emerged,
but region or coverage specific patterns were not evident
(Figure \ref{fig:hexpanel}).
Rather, we observed ME to be mostly positive across the GPO-LiDAR area,
and to be weakly related to the underlying distribution of FIA reference data.
We can also infer that large RMSE values were a reflection of extreme
individual outliers as they were often paired with reasonable MAE values.

Our choice to maintain FIA's probability sample in order to leverage unbiased
estimators of map agreement metrics came with some drawbacks.
Namely, we had to accept temporal lags of +/- 2 years between LiDAR acquisitions
and field inventories,
and we had to include structural zeros in our assessment dataset where FIA did
not record tree measurements in non-forest conditions.
It is likely that the former inflated our estimates of
error due to growth or disturbances occurring at plots between the time of
inventory and the time of LiDAR acquisition, though we have no means to
quantitatively confirm this possibility.
We can say with more certainty that the structural zeros introduced by
non-forest plots had a large impact on our map agreement assessment as evidenced
by the stack of non-zero predictions along the x-axis in plot to pixel
comparisons (Figure \ref{fig:riemann11})
and the large positive ME for plots classified as Grass/Shrub and Cropland
(Table \ref{tab:lcpriacc}).
Given the large non-zero biomass predictions produced at these plots, it is not
unreasonable to assume they contain trees unmeasured by FIA that would be
captured by LiDAR height metrics.
These structural zeros are likely the driving force behind the positive average
prediction error (ME) found in our assessments,
and may be the cause for the slight increase in Moran's I values computed with
the assessment dataset relative to those computed with the model dataset
(Supplementary Materials S6).

Nearly all of our assessments,
including performance against the testing partition of the model data
(Figure \ref{fig:modelperfscatter}),
the Riemann analysis (Figure \ref{fig:riemann11}),
agreement with the Menlove and Healey FIA estimates
(Figure \ref{fig:standardfia}),
and the weak relationship between ME and average FIA AGB estimate
(Figure \ref{fig:hexpanel} g)
reinforced our model's tendency towards lower accuracy at the extremes of
the FIA reference dataset.
These discrepancies can likely be attributed to the model structure and
training approach,
the aforementioned structural zeros in our map assessment dataset,
as well as the saturation problem inherent in LiDAR-AGB modeling
(St‐Onge, Hu, and Vega 2008) where models fail to predict the largest AGB values in the data
set.
In general, with the extremes of the response distribution occurring less often,
most models will be more accurate making predictions near the mean.

We also recognize the presence of uncertainty in the AGB reference data due to
allometric, measurement, and locational errors, as well as our growth adjustment
procedure used to boost the size of the model dataset,
though quantifying the magnitude of this uncertainty was outside the scope of
this paper.
Duncanson et al. (2017) indicated that different choices in allometric models
used to predict AGB from tree diameter and height measurements
can result in large variation (up to 20\%) of plot-level AGB estimates.
An improvement would be to embed measurement errors, as well as
allometric and growth adjustment uncertainty in a quantification of model
precision or uncertainty at the pixel level (CEOS 2021).
Such information could help to develop variance estimates for
any aggregation of pixel predictions that could be used in estimating AGB
in areas that lack field inventories (Dettmann et al. 2022; CEOS 2021).

\hypertarget{map-applications}{%
\subsection{Map Applications}\label{map-applications}}

Our rigorously evaluated map products have a range of applications
where knowledge of the spatial patterns of forest biomass (and by extension,
forest carbon pools) is needed for monitoring, reporting, and verification
efforts alongside policy or regulatory decision support.
These uses include the identification of forested areas for future monitoring,
protection, or management,
and for providing AGB as a predictor in subsequent ecological models.
Our AGB maps can also be leveraged as diverse training data
for models driven by spaceborne remote sensing platforms with more contiguous
spatial and temporal coverage than LiDAR patchworks,
providing the basis for landscape-scale carbon accounting
(Hudak et al. 2020; CEOS 2021).

\hypertarget{conclusion}{%
\section{Conclusion}\label{conclusion}}

Accurate AGB predictions at fine resolutions can provide landowners
and decision makers with valuable information on landscape patterns needed
to implement forest-based climate solutions, including reforestation
opportunities, avoided deforestation, and improved management for
carbon storage and sequestration.
We implemented a model-based approach leveraging extensive field inventory data
(FIA) and publicly available LiDAR coverages to develop AGB maps across NYS,
where forests are expected to contribute substantially as carbon sinks
towards achieving a net-zero carbon economy by 2050.
Although LiDAR point clouds provide detailed information on forest structure
that can yield superior models of forest biomass,
their limited coverage in both spatial and temporal domains produces
patchworks of disparate datasets over broad scales.
Our modeling approach,
and the comprehensive set of assessments demonstrated here,
addressed several of the common challenges inherent in using LiDAR patchworks
for AGB mapping, including a lack of temporally matching reference data and
data discrepancies among component LiDAR coverages.
Our results show that our approach and the resulting map products provide
accurate AGB information at scales relevant to forest and climate stewardship
in NYS.

\hypertarget{acknowledgements}{%
\section{Acknowledgements}\label{acknowledgements}}

We would like to thank the USDA FIA program for their data sharing and
cooperation, the NYS GPO for compiling LiDAR data and sharing tax-parcel data,
and the NYS Department of Environmental Conservation, Office of Climate Change,
for funding support.

\hypertarget{references}{%
\section*{References}\label{references}}
\addcontentsline{toc}{section}{References}

\hypertarget{refs}{}
\begin{CSLReferences}{1}{0}
\leavevmode\vadjust pre{\hypertarget{ref-Anderson2013}{}}%
Anderson, Ryan S., and Paul V. Bolstad. 2013. {``{Estimating Aboveground Biomass and Average Annual Wood Biomass Increment with Airborne Leaf-on and Leaf-off LiDAR in Great Lakes Forest Types}.''} \emph{Northern Journal of Applied Forestry} 30 (1): 16--22. \url{https://doi.org/10.5849/njaf.12-015}.

\leavevmode\vadjust pre{\hypertarget{ref-Ayrey2021}{}}%
Ayrey, Elias, Daniel J. Hayes, John B. Kilbride, Shawn Fraver, John A. Kershaw, Bruce D. Cook, and Aaron R. Weiskittel. 2021. {``Synthesizing Disparate LiDAR and Satellite Datasets Through Deep Learning to Generate Wall-to-Wall Regional Inventories for the Complex, Mixed-Species Forests of the Eastern United States.''} \emph{Remote Sensing} 13 (24). \url{https://doi.org/10.3390/rs13245113}.

\leavevmode\vadjust pre{\hypertarget{ref-Bechtold2005}{}}%
Bechtold, William A, and Paul L Patterson. 2005. \emph{The Enhanced Forest Inventory and Analysis Program--National Sampling Design and Estimation Procedures}. Vol. 80. USDA Forest Service, Southern Research Station. \url{https://doi.org/10.2737/SRS-GTR-80}.

\leavevmode\vadjust pre{\hypertarget{ref-Bivand2013}{}}%
Bivand, Roger S., Edzer Pebesma, and Virgilio Gomez-Rubio. 2013. \emph{{Applied Spatial Data Analysis with {R}, Second Edition}}. Springer, NY. \url{https://doi.org/10.1007/978-0-387-78171-6}.

\leavevmode\vadjust pre{\hypertarget{ref-Breiman2001}{}}%
Breiman, Leo. 2001a. {``Random Forests.''} \emph{Machine Learning} 45 (1): 5--32. \url{https://doi.org/10.1023/A:1010933404324}.

\leavevmode\vadjust pre{\hypertarget{ref-Breiman2Cultures}{}}%
---------. 2001b. {``Statistical Modeling: The Two Cultures.''} \emph{Statistical Science} 16 (3): 199--231. \url{https://doi.org/10.1214/ss/1009213726}.

\leavevmode\vadjust pre{\hypertarget{ref-Brown2020}{}}%
Brown, Jesslyn F., Heather J. Tollerud, Christopher P. Barber, Qiang Zhou, John L. Dwyer, James E. Vogelmann, Thomas R. Loveland, et al. 2020. {``{Lessons learned implementing an operational continuous United States national land change monitoring capability: The Land Change Monitoring, Assessment, and Projection (LCMAP) approach}.''} \emph{Remote Sensing of Environment} 238: 111356. \url{https://doi.org/10.1016/j.rse.2019.111356}.

\leavevmode\vadjust pre{\hypertarget{ref-IPCC2019}{}}%
Buendia, E, K Tanabe, A Kranjc, J Baasansuren, M Fukuda, S Ngarize, A Osako, Y Pyrozhenko, P Shermanau, and S Federici. 2019. {``Refinement to the 2006 IPCC Guidelines for National Greenhouse Gas Inventories.''} \emph{IPCC: Geneva, Switzerland} 5: 194.

\leavevmode\vadjust pre{\hypertarget{ref-CEOS}{}}%
CEOS. 2021. {``{Aboveground Woody Biomass Product Validation Good Practices Protocol}.''} \url{https://doi.org/10.5067/DOC/CEOSWGCV/LPV/AGB.001}.

\leavevmode\vadjust pre{\hypertarget{ref-Chazdon2016}{}}%
Chazdon, Robin L, Pedro HS Brancalion, Lars Laestadius, Aoife Bennett-Curry, Kathleen Buckingham, Chetan Kumar, Julian Moll-Rocek, Ima Célia Guimarães Vieira, and Sarah Jane Wilson. 2016. {``When Is a Forest a Forest? Forest Concepts and Definitions in the Era of Forest and Landscape Restoration.''} \emph{Ambio} 45 (5): 538--50. \url{https://doi.org/10.1007/s13280-016-0772-y}.

\leavevmode\vadjust pre{\hypertarget{ref-Chen2016}{}}%
Chen, Qi, and Ronald McRoberts. 2016. {``Statewide Mapping and Estimation of Vegetation Aboveground Biomass Using Airborne Lidar.''} In \emph{2016 {IEEE} International Geoscience and Remote Sensing Symposium ({IGARSS})}. {IEEE}. \url{https://doi.org/10.1109/igarss.2016.7730157}.

\leavevmode\vadjust pre{\hypertarget{ref-Cortes1995}{}}%
Cortes, Corinna, and Vladimir Vapnik. 1995. {``Support-Vector Networks.''} \emph{Machine Learning} 20 (3): 273--97. \url{https://doi.org/10.1007/bf00994018}.

\leavevmode\vadjust pre{\hypertarget{ref-Baston2021}{}}%
Daniel Baston. 2021. \emph{Exactextractr: Fast Extraction from Raster Datasets Using Polygons}. \url{https://CRAN.R-project.org/package=exactextractr}.

\leavevmode\vadjust pre{\hypertarget{ref-Dettmann2022}{}}%
Dettmann, Garret T, Philip J Radtke, John W Coulston, P Corey Green, Barry T Wilson, and Gretchen G Moisen. 2022. {``Review and Synthesis of Estimation Strategies to Meet Small Area Needs in Forest Inventory.''} \href{https://doi.org/10.3389/ffgc.2022.813569\%20\%20}{https://doi.org/10.3389/ffgc.2022.813569 }.

\leavevmode\vadjust pre{\hypertarget{ref-Duncanson2017}{}}%
Duncanson, Laura, Wenli Huang, Kristofer Johnson, Anu Swatantran, Ronald E McRoberts, and Ralph Dubayah. 2017. {``Implications of Allometric Model Selection for County-Level Biomass Mapping.''} \emph{Carbon Balance and Management} 12 (1): 1--11. \url{https://doi.org/10.1186/s13021-017-0086-9}.

\leavevmode\vadjust pre{\hypertarget{ref-Efron2020}{}}%
Efron, Bradley. 2020. {``Prediction, Estimation, and Attribution.''} \emph{Journal of the American Statistical Association} 115 (530): 636--55. \url{https://doi.org/10.1080/01621459.2020.1762613}.

\leavevmode\vadjust pre{\hypertarget{ref-Fassnacht2014}{}}%
Fassnacht, F. E., F. Hartig, H. Latifi, C. Berger, J. Hernández, P. Corvalán, and B. Koch. 2014. {``Importance of Sample Size, Data Type and Prediction Method for Remote Sensing-Based Estimations of Aboveground Forest Biomass.''} \emph{Remote Sensing of Environment} 154: 102--14. \url{https://doi.org/10.1016/j.rse.2014.07.028}.

\leavevmode\vadjust pre{\hypertarget{ref-Friedman2002}{}}%
Friedman, Jerome H. 2002. {``Stochastic Gradient Boosting.''} \emph{Computational Statistics and Data Analysis} 38 (4): 367--78. \url{https://doi.org/10.1016/S0167-9473(01)00065-2}.

\leavevmode\vadjust pre{\hypertarget{ref-Fritz2017}{}}%
Fritz, Steffen, Linda See, Christoph Perger, Ian McCallum, Christian Schill, Dmitry Schepaschenko, Martina Duerauer, et al. 2017. {``A Global Dataset of Crowdsourced Land Cover and Land Use Reference Data.''} \emph{Scientific Data} 4 (1). \url{https://doi.org/10.1038/sdata.2017.75}.

\leavevmode\vadjust pre{\hypertarget{ref-Gobakken2008}{}}%
Gobakken, Terje, and Erik Næsset. 2008. {``Assessing Effects of Laser Point Density, Ground Sampling Intensity, and Field Sample Plot Size on Biophysical Stand Properties Derived from Airborne Laser Scanner Data.''} \emph{Canadian Journal of Forest Research} 38 (5): 1095--1109. \url{https://doi.org/10.1139/X07-219}.

\leavevmode\vadjust pre{\hypertarget{ref-Goncalves2017}{}}%
Gonçalves, Fabio, Robert Treuhaft, Beverly Law, André Almeida, Wayne Walker, Alessandro Baccini, João dos Santos, and Paulo Graça. 2017. {``Estimating Aboveground Biomass in Tropical Forests: Field Methods and Error Analysis for the Calibration of Remote Sensing Observations.''} \emph{Remote Sensing} 9 (1): 47. \url{https://doi.org/10.3390/rs9010047}.

\leavevmode\vadjust pre{\hypertarget{ref-Gray2012}{}}%
Gray, Andrew N, Thomas J Brandeis, John D Shaw, William H McWilliams, and Patrick Miles. 2012. {``{Forest Inventory and Analysis Database of the United States of America (FIA)}.''} \emph{Biodiversity and Ecology} 4: 225--31. \url{https://doi.org/10.7809/b-e.00079}.

\leavevmode\vadjust pre{\hypertarget{ref-Hauglin2021}{}}%
Hauglin, Marius, Johannes Rahlf, Johannes Schumacher, Rasmus Astrup, and Johannes Breidenbach. 2021. {``Large Scale Mapping of Forest Attributes Using Heterogeneous Sets of Airborne Laser Scanning and National Forest Inventory Data.''} \emph{Forest Ecosystems} 8 (1). \url{https://doi.org/10.1186/s40663-021-00338-4}.

\leavevmode\vadjust pre{\hypertarget{ref-Hawbaker2010}{}}%
Hawbaker, Todd J., Terje Gobakken, Adrian Lesak, Eric Trømborg, Kirk Contrucci, and Volker Radeloff. 2010. {``{Light Detection and Ranging-Based Measures of Mixed Hardwood Forest Structure}.''} \emph{Forest Science} 56 (3): 313--26. \url{https://doi.org/10.1093/forestscience/56.3.313}.

\leavevmode\vadjust pre{\hypertarget{ref-Raster2021}{}}%
Hijmans, Robert J. 2021. \emph{Raster: Geographic Data Analysis and Modeling}. \url{https://CRAN.R-project.org/package=raster}.

\leavevmode\vadjust pre{\hypertarget{ref-terra}{}}%
---------. 2022. \emph{Terra: Spatial Data Analysis}. \url{https://CRAN.R-project.org/package=terra}.

\leavevmode\vadjust pre{\hypertarget{ref-Hoppus2005}{}}%
Hoppus, Michael, and Andrew Lister. 2005. {``The Status of Accurately Locating Forest Inventory and Analysis Plots Using the Global Positioning System.''} In \emph{Proceedings of the Seventh Annual Forest Inventory and Analysis Symposium}. \url{https://www.nrs.fs.fed.us/pubs/7040}.

\leavevmode\vadjust pre{\hypertarget{ref-Houghton2012}{}}%
Houghton, R. A., J. I. House, J. Pongratz, G. R. van der Werf, R. S. DeFries, M. C. Hansen, C. Le Quéré, and N. Ramankutty. 2012. {``Carbon Emissions from Land Use and Land-Cover Change.''} \emph{Biogeosciences} 9 (12): 5125--42. \url{https://doi.org/10.5194/bg-9-5125-2012}.

\leavevmode\vadjust pre{\hypertarget{ref-Houghton2005}{}}%
Houghton, RA. 2005. {``Aboveground Forest Biomass and the Global Carbon Balance.''} \emph{Global Change Biology} 11 (6): 945--58.

\leavevmode\vadjust pre{\hypertarget{ref-Huang2019}{}}%
Huang, Wenli, Katelyn Dolan, Anu Swatantran, Kristofer Johnson, Hao Tang, Jarlath O'Neil-Dunne, Ralph Dubayah, and George Hurtt. 2019. {``{High-resolution mapping of aboveground biomass for forest carbon monitoring system in the Tri-State region of Maryland, Pennsylvania and Delaware, {USA}}.''} \emph{Environmental Research Letters} 14 (9): 095002. \url{https://doi.org/10.1088/1748-9326/ab2917}.

\leavevmode\vadjust pre{\hypertarget{ref-Hudak2020}{}}%
Hudak, Andrew T, Patrick A Fekety, Van R Kane, Robert E Kennedy, Steven K Filippelli, Michael J Falkowski, Wade T Tinkham, et al. 2020. {``A Carbon Monitoring System for Mapping Regional, Annual Aboveground Biomass Across the Northwestern {USA}.''} \emph{Environmental Research Letters} 15 (9): 095003. \url{https://doi.org/10.1088/1748-9326/ab93f9}.

\leavevmode\vadjust pre{\hypertarget{ref-Hurtt2019}{}}%
Hurtt, G, M Zhao, R Sahajpal, A Armstrong, R Birdsey, E Campbell, K Dolan, et al. 2019. {``{Beyond {MRV}: high-resolution forest carbon modeling for climate mitigation planning over Maryland, {USA}}.''} \emph{Environmental Research Letters} 14 (4): 045013. \url{https://doi.org/10.1088/1748-9326/ab0bbe}.

\leavevmode\vadjust pre{\hypertarget{ref-lidrRSE}{}}%
Jean-Romain, Roussel, David Auty, Nicholas C. Coops, Piotr Tompalski, Tristan R. H. Goodbody, Andrew Sánchez Meador, Jean-François Bourdon, Florian de Boissieu, and Alexis Achim. 2020. {``{lidR: An R package for analysis of Airborne Laser Scanning (ALS) data}.''} \emph{Remote Sensing of Environment} 251: 112061. \url{https://doi.org/10.1016/j.rse.2020.112061}.

\leavevmode\vadjust pre{\hypertarget{ref-Johnson2015}{}}%
Johnson, Kristofer D, Richard Birdsey, Jason Cole, Anu Swatantran, Jarlath O'Neil-Dunne, Ralph Dubayah, and Andrew Lister. 2015. {``Integrating LIDAR and Forest Inventories to Fill the Trees Outside Forests Data Gap.''} \emph{Environmental Monitoring and Assessment} 187 (10): 1--8. \url{https://doi.org/10.1007/s10661-015-4839-1}.

\leavevmode\vadjust pre{\hypertarget{ref-Johnson2014}{}}%
Johnson, Kristofer D, Richard Birdsey, Andrew O Finley, Anu Swantaran, Ralph Dubayah, Craig Wayson, and Rachel Riemann. 2014. {``Integrating Forest Inventory and Analysis Data into a {LIDAR}-Based Carbon Monitoring System.''} \emph{Carbon Balance and Management} 9 (1). \url{https://doi.org/10.1186/1750-0680-9-3}.

\leavevmode\vadjust pre{\hypertarget{ref-Alexandros2004}{}}%
Karatzoglou, Alexandros, Alexandros Smola, Kurt Hornik, and Achim Zeileis. 2004. {``{kernlab - An S4 Package for Kernel Methods in R}.''} \emph{Journal of Statistical Software, Articles} 11 (9): 1--20. \url{https://doi.org/10.18637/jss.v011.i09}.

\leavevmode\vadjust pre{\hypertarget{ref-Guolin2017}{}}%
Ke, Guolin, Qi Meng, Thomas Finley, Taifeng Wang, Wei Chen, Weidong Ma, Qiwei Ye, and Tie-Yan Liu. 2017. {``{LightGBM: A Highly Efficient Gradient Boosting Decision Tree}.''} In \emph{Advances in Neural Information Processing Systems}, edited by I. Guyon, U. V. Luxburg, S. Bengio, H. Wallach, R. Fergus, S. Vishwanathan, and R. Garnett. Vol. 30. Curran Associates, Inc. \url{https://proceedings.neurips.cc/paper/2017/file/6449f44a102fde848669bdd9eb6b76fa-Paper.pdf}.

\leavevmode\vadjust pre{\hypertarget{ref-Guolin2021}{}}%
Ke, Guolin, Damien Soukhavong, James Lamb, Qi Meng, Thomas Finley, Taifeng Wang, Wei Chen, Weidong Ma, Qiwei Ye, and Tie-Yan Liu. 2021. \emph{Lightgbm: Light Gradient Boosting Machine}. \url{https://CRAN.R-project.org/package=lightgbm}.

\leavevmode\vadjust pre{\hypertarget{ref-Kennedy2018}{}}%
Kennedy, Robert E, Janet Ohmann, Matt Gregory, Heather Roberts, Zhiqiang Yang, David M Bell, Van Kane, et al. 2018. {``An Empirical, Integrated Forest Biomass Monitoring System.''} \emph{Environmental Research Letters} 13 (2): 025004. \url{https://doi.org/10.1088/1748-9326/aa9d9e}.

\leavevmode\vadjust pre{\hypertarget{ref-Lroe2013}{}}%
L'Roe, Andrew W, and Shorna Broussard Allred. 2013. {``{Thriving or Surviving? Forester Responses to Private Forestland Parcelization in New York State}.''} \emph{Small-Scale Forestry} 12 (3): 353--76. \url{https://doi.org/10.1007/s11842-012-9216-0}.

\leavevmode\vadjust pre{\hypertarget{ref-Lu2014}{}}%
Lu, Dengsheng, Qi Chen, Guangxing Wang, Lijuan Liu, Guiying Li, and Emilio Moran. 2014. {``A Survey of Remote Sensing-Based Aboveground Biomass Estimation Methods in Forest Ecosystems.''} \emph{International Journal of Digital Earth} 9 (1): 63--105. \url{https://doi.org/10.1080/17538947.2014.990526}.

\leavevmode\vadjust pre{\hypertarget{ref-Shrubland}{}}%
Mahoney, Michael J, Lucas K Johnson, and Colin M Beier. 2022. {``Classification and Mapping of Low-Statured 'Shrubland' Cover Types in Post-Agricultural Landscapes of the US Northeast.''} arXiv. \url{https://doi.org/10.48550/ARXIV.2205.05047}.

\leavevmode\vadjust pre{\hypertarget{ref-McRoberts2011}{}}%
McRoberts, Ronald E. 2011. {``Satellite Image-Based Maps: Scientific Inference or Pretty Pictures?''} \emph{Remote Sensing of Environment} 115 (2): 715--24. \url{https://doi.org/10.1016/j.rse.2010.10.013}.

\leavevmode\vadjust pre{\hypertarget{ref-Mcroberts2018b}{}}%
McRoberts, Ronald E., Qi Chen, Brian F. Walters, and Daniel J. Kaisershot. 2018. {``The Effects of Global Positioning System Receiver Accuracy on Airborne Laser Scanning-Assisted Estimates of Aboveground Biomass.''} \emph{Remote Sensing of Environment} 207: 42--49. \url{https://doi.org/10.1016/j.rse.2017.09.036}.

\leavevmode\vadjust pre{\hypertarget{ref-Menlove2020}{}}%
Menlove, James, and Sean P. Healey. 2020. {``{A Comprehensive Forest Biomass Dataset for the {USA} Allows Customized Validation of Remotely Sensed Biomass Estimates}.''} \emph{Remote Sensing} 12 (24): 4141. \url{https://doi.org/10.3390/rs12244141}.

\leavevmode\vadjust pre{\hypertarget{ref-Meyer2021}{}}%
Meyer, Hanna, and Edzer Pebesma. 2021. {``{Predicting into unknown space? Estimating the area of applicability of spatial prediction models}.''} \emph{Methods in Ecology and Evolution} 12 (9): 1620--33. \url{https://doi.org/10.1111/2041-210x.13650}.

\leavevmode\vadjust pre{\hypertarget{ref-Moran1950}{}}%
Moran, Patrick AP. 1950. {``{Notes on Continuous Stochastic Phenomena}.''} \emph{Biometrika} 37 (1/2): 17--23. \url{https://doi.org/10.2307/2332142}.

\leavevmode\vadjust pre{\hypertarget{ref-Nilsson2017}{}}%
Nilsson, Mats, Karin Nordkvist, Jonas Jonzén, Nils Lindgren, Peder Axensten, Jörgen Wallerman, Mikael Egberth, et al. 2017. {``{A nationwide forest attribute map of Sweden predicted using airborne laser scanning data and field data from the National Forest Inventory}.''} \emph{Remote Sensing of Environment} 194: 447--54. https://doi.org/\url{https://doi.org/10.1016/j.rse.2016.10.022}.

\leavevmode\vadjust pre{\hypertarget{ref-sf2018}{}}%
Pebesma, Edzer. 2018. {``{Simple Features for R: Standardized Support for Spatial Vector Data}.''} \emph{{The R Journal}} 10 (1): 439--46. \url{https://doi.org/10.32614/RJ-2018-009}.

\leavevmode\vadjust pre{\hypertarget{ref-Pengra2020}{}}%
Pengra, Bruce, Steve Stehman, Josephine A Horton, Daryn J Dockter, Todd A Schroeder, Zhiqiang Yang, Alex J Hernandez, et al. 2020. {``LCMAP Reference Data Product 1984-2018 Land Cover, Land Use and Change Process Attributes (Ver. 1.2, November 2021).''} \url{https://doi.org/10.5066/P9ZWOXJ7}.

\leavevmode\vadjust pre{\hypertarget{ref-Pflugmacher2014}{}}%
Pflugmacher, Dirk, Warren B. Cohen, Robert E. Kennedy, and Zhiqiang Yang. 2014. {``{Using Landsat-derived disturbance and recovery history and lidar to map forest biomass dynamics}.''} \emph{Remote Sensing of Environment} 151: 124--37. \url{https://doi.org/10.1016/j.rse.2013.05.033}.

\leavevmode\vadjust pre{\hypertarget{ref-R}{}}%
R Core Team. 2021. \emph{R: A Language and Environment for Statistical Computing}. Vienna, Austria: R Foundation for Statistical Computing. \url{https://www.R-project.org/}.

\leavevmode\vadjust pre{\hypertarget{ref-Riemann2010}{}}%
Riemann, Rachel, Barry Tyler Wilson, Andrew Lister, and Sarah Parks. 2010. {``{An effective assessment protocol for continuous geospatial datasets of forest characteristics using {USFS} Forest Inventory and Analysis ({FIA}) data}.''} \emph{Remote Sensing of Environment} 114 (10): 2337--52. \url{https://doi.org/10.1016/j.rse.2010.05.010}.

\leavevmode\vadjust pre{\hypertarget{ref-lidrCRAN}{}}%
Roussel, Jean-Romain, and David Auty. 2020. \emph{Airborne LiDAR Data Manipulation and Visualization for Forestry Applications}. \url{https://cran.r-project.org/package=lidR}.

\leavevmode\vadjust pre{\hypertarget{ref-Skowronksi2012}{}}%
Skowronski, Nicholas S, and Andrew J Lister. 2012. {``{Utility of LiDAR for large area forest inventory applications}.''} In \emph{In: Morin, Randall s.; Liknes, Greg c., Comps. Moving from Status to Trends: Forest Inventory and Analysis (FIA) Symposium 2012; 2012 December 4-6; Baltimore, MD. Gen. Tech. Rep. NRS-p-105. Newtown Square, PA: US Department of Agriculture, Forest Service, Northern Research Station.{[}CD-ROM{]}: 410-413.}, 410--13. \url{https://www.fs.usda.gov/treesearch/pubs/42792}.

\leavevmode\vadjust pre{\hypertarget{ref-Stehman2019}{}}%
Stehman, Stephen V., and Giles M. Foody. 2019. {``Key Issues in Rigorous Accuracy Assessment of Land Cover Products''} 231 (September): 111199. \url{https://doi.org/10.1016/j.rse.2019.05.018}.

\leavevmode\vadjust pre{\hypertarget{ref-Stonge2008}{}}%
St‐Onge, B., Y. Hu, and C. Vega. 2008. {``{Mapping the height and above‐ground biomass of a mixed forest using lidar and stereo Ikonos images}.''} \emph{International Journal of Remote Sensing} 29 (5): 1277--94. \url{https://doi.org/10.1080/01431160701736505}.

\leavevmode\vadjust pre{\hypertarget{ref-Thompson2011}{}}%
Thompson, Jonathan R., David R. Foster, Robert Scheller, and David Kittredge. 2011. {``{The influence of land use and climate change on forest biomass and composition in Massachusetts, USA}.''} \emph{Ecological Applications} 21 (7): 2425--44. \url{https://doi.org/10.1890/10-2383.1}.

\leavevmode\vadjust pre{\hypertarget{ref-NASS}{}}%
USDA National Agricultural Statistics Service. 2019. {``{2017 Census of Agriculture}.''} \url{https://www.nass.usda.gov/Publications/AgCensus/2017/index.php}.

\leavevmode\vadjust pre{\hypertarget{ref-White2015}{}}%
White, Joanne C., John T. T. R. Arnett, Michael A. Wulder, Piotr Tompalski, and Nicholas C. Coops. 2015. {``Evaluating the Impact of Leaf-on and Leaf-Off Airborne Laser Scanning Data on the Estimation of Forest Inventory Attributes with the Area-Based Approach.''} \emph{Canadian Journal of Forest Research} 45 (11): 1498--1513. \url{https://doi.org/10.1139/cjfr-2015-0192}.

\leavevmode\vadjust pre{\hypertarget{ref-White2013}{}}%
White, Joanne C., Michael A. Wulder, Andrés Varhola, Mikko Vastaranta, Nicholas C. Coops, Bruce D. Cook, Doug Pitt, and Murray Woods. 2013. {``A Best Practices Guide for Generating Forest Inventory Attributes from Airborne Laser Scanning Data Using an Area-Based Approach.''} \emph{The Forestry Chronicle} 89 (06): 722--23. \url{https://doi.org/10.5558/tfc2013-132}.

\leavevmode\vadjust pre{\hypertarget{ref-Wintle2003}{}}%
Wintle, B. A., M. A. McCarthy, C. T. Volinksy, and R. P. Kavanagh. 2003. {``The Use of Bayesian Model Averaging to Better Represent Uncertainty in Ecological Models.''} \emph{Conservation Biology} 17 (6): 1579--90. \url{https://doi.org/10.1111/j.1523-1739.2003.00614.x}.

\leavevmode\vadjust pre{\hypertarget{ref-Wolpert1992}{}}%
Wolpert, David H. 1992. {``Stacked Generalization.''} \emph{Neural Networks} 5 (2): 241--59. \url{https://doi.org/10.1016/S0893-6080(05)80023-1}.

\leavevmode\vadjust pre{\hypertarget{ref-Woodall2015}{}}%
Woodall, Christopher W, John W Coulston, Grant M Domke, Brian F Walters, David N Wear, James E Smith, Hans-Erik Andersen, et al. 2015. {``The US Forest Carbon Accounting Framework: Stocks and Stock Change, 1990-2016.''} \emph{Gen. Tech. Rep. NRS-154. Newtown Square, PA: US Department of Agriculture, Forest Service, Northern Research Station. 49 p.} 154: 1--49. \url{https://doi.org/10.2737/NRS-GTR-154}.

\leavevmode\vadjust pre{\hypertarget{ref-Wright2017}{}}%
Wright, Marvin N., and Andreas Ziegler. 2017. {``Ranger: A Fast Implementation of Random Forests for High Dimensional Data in c++ and r.''} \emph{Journal of Statistical Software, Articles} 77 (1): 1--17. \url{https://doi.org/10.18637/jss.v077.i01}.

\leavevmode\vadjust pre{\hypertarget{ref-Landsat50}{}}%
Wulder, Michael A., David P. Roy, Volker C. Radeloff, Thomas R. Loveland, Martha C. Anderson, David M. Johnson, Sean Healey, et al. 2022. {``Fifty Years of Landsat Science and Impacts.''} \emph{Remote Sensing of Environment} 280: 113195. \url{https://doi.org/10.1016/j.rse.2022.113195}.

\leavevmode\vadjust pre{\hypertarget{ref-Yang2018}{}}%
Yang, Limin, Suming Jin, Patrick Danielson, Collin Homer, Leila Gass, Stacie M. Bender, Adam Case, et al. 2018. {``{A new generation of the United States National Land Cover Database: Requirements, research priorities, design, and implementation strategies}.''} \emph{{ISPRS} Journal of Photogrammetry and Remote Sensing} 146 (December): 108--23. \url{https://doi.org/10.1016/j.isprsjprs.2018.09.006}.

\leavevmode\vadjust pre{\hypertarget{ref-Zhu2014}{}}%
Zhu, Zhe, and Curtis E. Woodcock. 2014. {``{Continuous change detection and classification of land cover using all available Landsat data}.''} \emph{Remote Sensing of Environment} 144: 152--71. \url{https://doi.org/10.1016/j.rse.2014.01.011}.

\end{CSLReferences}

\end{document}


\maketitle

{
\setcounter{tocdepth}{2}
\tableofcontents
}
\clearpage

\hypertarget{s1-modeling-predictors}{%
\section{S1: Modeling Predictors}\label{s1-modeling-predictors}}

\begin{table}[!h]

\caption{\label{tab:predictors}Definitions of predictors used for model fitting.}
\centering
\fontsize{10}{12}\selectfont
\begin{tabular}[t]{>{\raggedright\arraybackslash}p{10em}>{\raggedright\arraybackslash}p{22em}l}
\toprule
\multicolumn{1}{c}{Predictor} & \multicolumn{1}{c}{Definition} & \multicolumn{1}{c}{Group}\\
\midrule
H0, H10, ... H100, H95, H99 & Decile heights of returns, in meters, as well as 95th and 99th percentile return heights. & LiDAR\\
\addlinespace
D10, D20... D90 & Density of returns above a certain height, as a proportion. After return height is divided into 10 equal bins ranging from 0 to the maximum height of returns, this value reflects the proportion of returns at or above each breakpoint. & LiDAR\\
\addlinespace
ZMEAN, ZMEAN\_C & Mean height of all returns (ZMEAN) and all returns above 2.5m (ZMEAN\_C) & LiDAR\\
\addlinespace
Z\_KURT, Z\_SKEW & Kurtosis and skewness of height of all returns & LiDAR\\
\addlinespace
QUAD\_MEAN, QUAD\_MEAN\_C & Quadratic mean height of all returns (QUAD\_MEAN) and all returns above 2.5m (QUAD\_MEAN\_C) & LiDAR\\
\addlinespace
CV, CV\_C & Coefficient of variation for heights of all returns (CV) and all returns above 2.5m (CV\_C) & LiDAR\\
\addlinespace
L2, L3, L4, L\_CV, L\_SKEW, L\_KURT & L-moments and their ratios as defined by Hosking (1990), calculated for heights of all returns & LiDAR\\
\addlinespace
CANCOV & Ratio of returns above 2.5m to all returns (Pflugmacher et al. 2012) & LiDAR\\
\addlinespace
HVOL & CANCOV * ZMEAN (Pflugmacher et al. 2012) & LiDAR\\
\addlinespace
RPC1 & Ratio of first returns to all returns (Pflugmacher et al. 2012) & LiDAR\\
\addlinespace
TMIN, TMIN, PRECIP & Climate variables in the form of 30 year normals (1980-2010) for minimum and maximum annual temperature (C) and average annual precipitation (mm) from Daly (2008) & Climate\\
\addlinespace
ELEV, SLOPE, ASPECT, TWI & Topographic variables computed from a 30m DEM downloaded using the terrainr package (Mahoney 2021) & Topography\\
\addlinespace
TAX\_CODE\_* & Individual tax code classifications as defined by NYS Department of Taxation and Finances (2019) & Tax\\
\addlinespace
TAX\_CATEGORY\_* & Broad grouping of tax codes as defined by NYS Department of Taxation and Finances (2019) & Tax\\
\bottomrule
\end{tabular}
\end{table}

\newpage{}

\hypertarget{s2-tax-parcel-data}{%
\section{S2: Tax Parcel Data}\label{s2-tax-parcel-data}}

A shapefile containing tax parcels and associated tax parcel codes was obtained
under agreement with the NYS GPO.
This shapefile was rasterized using to grid geometry matching all other
predictor variables using the ``fasterize'' R package (Ross 2020).
No-data pixels were filled with the default value of 1000,
and special NYC tax parcel codes were remapped to the default value of 2000.
Codes (Table \ref{tab:codes}) were mapped to categories
(Table \ref{tab:categories}) representing more broad groupings of similar
use and ownership types.
Codes representing \textless{} 1 \% of plots and categories representing \textless{} 5\% of plots
in the calibration dataset were ignored.
Codes or categories with less representation than these defined thresholds
cause issues with model fitting for the same reason as any near-0 variance
variable.
Further descriptions of these variables can be found on the NYS department of
taxation website:
\url{https://www.tax.ny.gov/research/property/assess/manuals/prclas.htm}.

\clearpage

\begin{longtable}[t]{>{\raggedright\arraybackslash}p{10em}>{\raggedright\arraybackslash}p{10em}>{\raggedright\arraybackslash}p{20em}}
\caption{\label{tab:codes}Tax Code Indicator Variables}\\
\toprule
\multicolumn{1}{c}{Variable Name} & \multicolumn{1}{c}{Category} & \multicolumn{1}{c}{Description}\\
\midrule
\endfirsthead
\caption[]{\label{tab:codes}Tax Code Indicator Variables \textit{(continued)}}\\
\toprule
\multicolumn{1}{c}{Variable Name} & \multicolumn{1}{c}{Category} & \multicolumn{1}{c}{Description}\\
\midrule
\endhead

\endfoot
\bottomrule
\endlastfoot
TAX\_CODE\_105 & Agricultural & Agricultural vacant land (productive)\\
\addlinespace
TAX\_CODE\_112 & Agricultural & Dairy products: milk, butter and cheese\\
\addlinespace
TAX\_CODE\_120 & Agricultural & Field crops\\
\addlinespace
TAX\_CODE\_210 & Residential & One family year-round residence\\
\addlinespace
TAX\_CODE\_240 & Residential & Rural residence with acreage\\
\addlinespace
TAX\_CODE\_241 & Residential & Primary residential, also used in agricultural production\\
\addlinespace
TAX\_CODE\_260 & Residential & Seasonal residences\\
\addlinespace
TAX\_CODE\_270 & Residential & Mobile home\\
\addlinespace
TAX\_CODE\_280 & Residential & Residential\\
\addlinespace
TAX\_CODE\_312 & Vacant land & Residential land including a small improvement (not used for living accommodations)\\
\addlinespace
TAX\_CODE\_314 & Vacant land & Rural vacant lots of 10 acres or less\\
\addlinespace
TAX\_CODE\_322 & Vacant land & Residential vacant land over 10 acres\\
\addlinespace
TAX\_CODE\_323 & Vacant land & Other rural vacant lands\\
\addlinespace
TAX\_CODE\_910 & Wild, forested, conservation lands and public parks & Private wild and forest lands except for private hunting and fishing clubs\\
\addlinespace
TAX\_CODE\_911 & Wild, forested, conservation lands and public parks & Forest land under Section 480 of the Real Property Tax Law\\
\addlinespace
TAX\_CODE\_912 & Wild, forested, conservation lands and public parks & Forest land under Section 480-a of the Real Property Tax Law\\
\addlinespace
TAX\_CODE\_930 & Wild, forested, conservation lands and public parks & State-owned forest lands\\
\addlinespace
TAX\_CODE\_931 & Wild, forested, conservation lands and public parks & State-owned land (forest Preserve) in the Adirondack or Catskill Parks taxable under Section 532-a of the Real Property Tax Law\\
\addlinespace
TAX\_CODE\_932 & Wild, forested, conservation lands and public parks & State-owned land other than forest preserve covered under Section 532-b, c, d, e, f, or g of the Real Property Tax Law\\
\addlinespace
TAX\_CODE\_941 & Wild, forested, conservation lands and public parks & State-owned reforested land taxable under Sections 534 and 536 of the Real Property Tax Law\\
\addlinespace
TAX\_CODE\_1000 & Missing data & No tax code present - this value represents missing data.\\
\addlinespace
TAX\_CODE\_2000 & NYC & Special tax codes for NYC were remapped to this default value.\\*
\end{longtable}

\begin{table}

\caption{\label{tab:categories}Tax Category Indicator Variables}
\centering
\begin{tabular}[t]{>{\raggedright\arraybackslash}p{15em}>{\raggedright\arraybackslash}p{10em}>{\raggedright\arraybackslash}p{15em}}
\toprule
\multicolumn{1}{c}{Variable Name} & \multicolumn{1}{c}{Category} & \multicolumn{1}{c}{Description}\\
\midrule
TAX\_CATEGEORY\_100 & Agricultural & Property used for the production of crops or livestock.\\
\addlinespace
TAX\_CATEGEORY\_200 & Residential & Property used for human habitation. Living accommodations such as hotels, motels, and apartments are in the Commercial category - 400.\\
\addlinespace
TAX\_CATEGEORY\_300 & Vacant land & Property that is not in use, is in temporary use, or lacks permanent improvement.\\
\addlinespace
TAX\_CATEGEORY\_900 & Wild, forested, conservation lands and public parks & Reforested lands, preserves, and private hunting and fishing clubs.\\
\bottomrule
\end{tabular}
\end{table}

\clearpage
\newpage{}

\hypertarget{model}{%
\section{S3: Model Validation}\label{model}}

\begin{figure}
\includegraphics[width=1\linewidth]{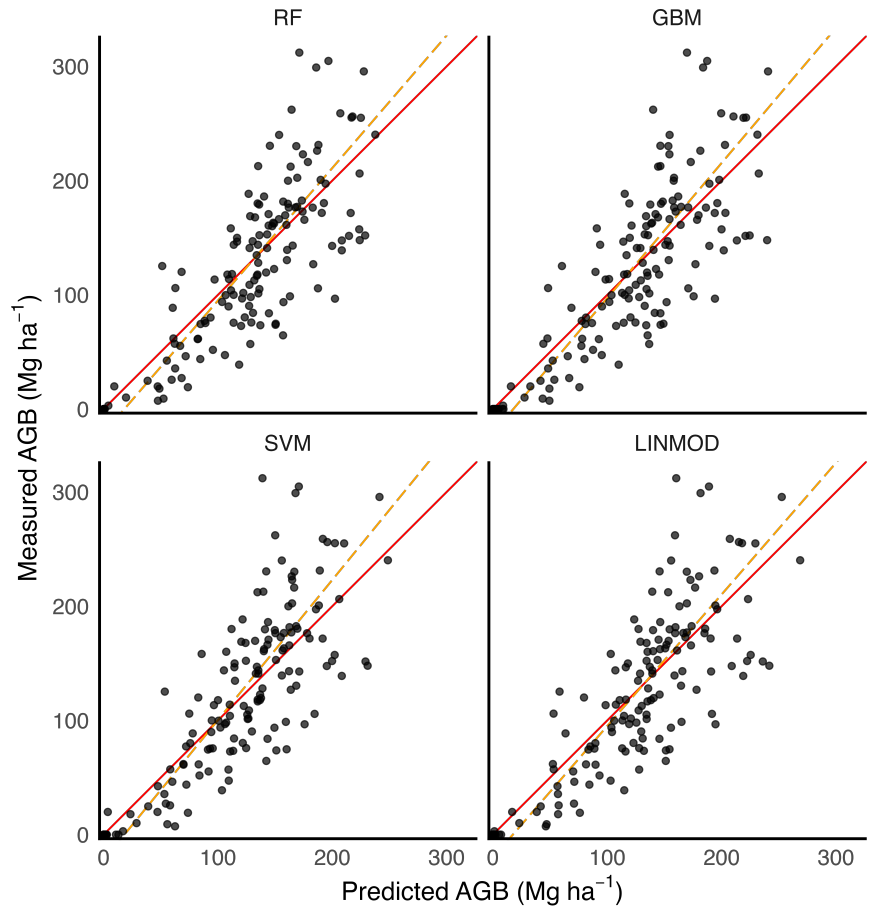} \caption{Measured vs predicted AGB scatter plot for four component models against the testing portion of the model dataset. AGB values in Mg ha$^{-1}$. GMFR trend line shown with dashed (orange) line, and 1:1 line shown with solid (red) line}\label{fig:modelperfscatter}
\end{figure}

Each of the three component models (RF, SVM, GBM) and the ``stacked''
linear regression ensemble model (LINMOD) were assessed against the 20\% holdout
partition (Table \ref{tab:modelperf} and Figure \ref{fig:modelperfscatter}).

\begin{table}[H]

\caption{\label{tab:modelperf}Model performance metrics (as defined in section S3) against 20\% holdout partition (n = 171; RMSE, ME in Mgha$^{-1}$).}
\centering
\begin{tabular}[t]{lrrrr}
\toprule
\multicolumn{1}{c}{ } & \multicolumn{1}{c}{RF} & \multicolumn{1}{c}{GBM} & \multicolumn{1}{c}{SVM} & \multicolumn{1}{c}{LINMOD}\\
\midrule
RMSE & 40.55 & 41.02 & 41.63 & 41.07\\
\addlinespace
\% RMSE & 37.34 & 37.78 & 38.34 & 37.82\\
\addlinespace
MAE & 29.08 & 29.68 & 29.27 & 29.36\\
\addlinespace
ME & 3.14 & 1.05 & -1.90 & 3.10\\
\addlinespace
R$^2$ & 0.75 & 0.74 & 0.73 & 0.74\\
\bottomrule
\end{tabular}
\end{table}

\newpage{}

\hypertarget{s4-area-of-applicability}{%
\section{S4: Area of Applicability}\label{s4-area-of-applicability}}

Following Meyer and Pebesma (2021) we computed an area of applicability (AOA) surface
to identify individual pixels whose predictor data is beyond a threshold
predictor-space distance from the training data.
The predictor variables are scaled by dividing mean-centered values by their
respective standard deviations and then weighted by their importance to the
linear ensemble model.
The distance for a given observation or pixel is computed as the minimum
Euclidean distance (in predictor space) from said observation to the
training data observations.
Using this distance, a dissimilarity index (DI) distance is then standardized,\\
by dividing by the average distances in the training data.
A threshold to identify ``in'' observations and ``out'' observations for the AOA is
defined as the 75th percentile plus 1.5 times the IQR of the DI values in the
20\% testing partition of the calibration data.
Observations with computed DIs beyond this threshold are identified as
outside the AOA.

\hypertarget{predictor-weights}{%
\subsubsection{Predictor Weights}\label{predictor-weights}}

Variable importance measures are required to weight individual predictor
contributions to the DI distance by their relative importance to the linear
ensemble model's predictions.
Without predictor weights, a single noisy predictor that has little impact
on the model's prediction might cause a pixel or observation to be deemed
``out'' of the AOA.
Variable importance was computed for the linear ensemble model using a
permutation approach (Breiman 2001).
Since many of our predictors are highly correlated, we first
clustered them based on spearman rank correlation Euclidean distances.
A cluster dendrogram was used to identify seven distinct variable clusters.
For each p predictor, a separate linear ensemble model is trained.
The pth predictor is selected, and all other predictors in the pth predictor's
cluster are excluded.
A single predictor from the remaining 6 clusters is selected.
The values for the pth predictor are then shuffled, breaking any potential
relationship between the pth predictor and the response (AGB).
The model is then trained with these seven predictors
(following the approach in Section S3), and RMSE is assessed
via LOO-cross-validation on the 80\% training partition.
Variable importance is then computed as the delta RMSE from the pth predictor
model against the full LIMMOD model trained with all p predictors.
The variable importances are then standardized to produce weights, such that the
sum of all p weights is equal to 1.
In this analysis default hyperparameter values are used for all component
ML models.

\newpage{}

\hypertarget{s5-map-agreement-assessment}{%
\section{S5: Map Agreement Assessment}\label{s5-map-agreement-assessment}}

Following the Riemann sampling intensity correction,
all comparisons were developed by extracting the area-weighted mean of mapped
pixels intersecting each FIA plot (Riemann et al. 2010).
LCMAP or AOA masked pixels were excluded from the map agreement assessment such
that any FIA plot intersecting a masked pixel was excluded from the analysis.
The hexagons were tessellated randomly across the GPO-LiDAR region with no
regard to FIA sample grid alignment.
The geometric mean functional relationship (GMFR) slope and intercepts were
computed following the Riemann approach
(which follows Draper and Smith (1998) and Ricker (1984)).

\begin{figure}
\includegraphics[width=1\linewidth]{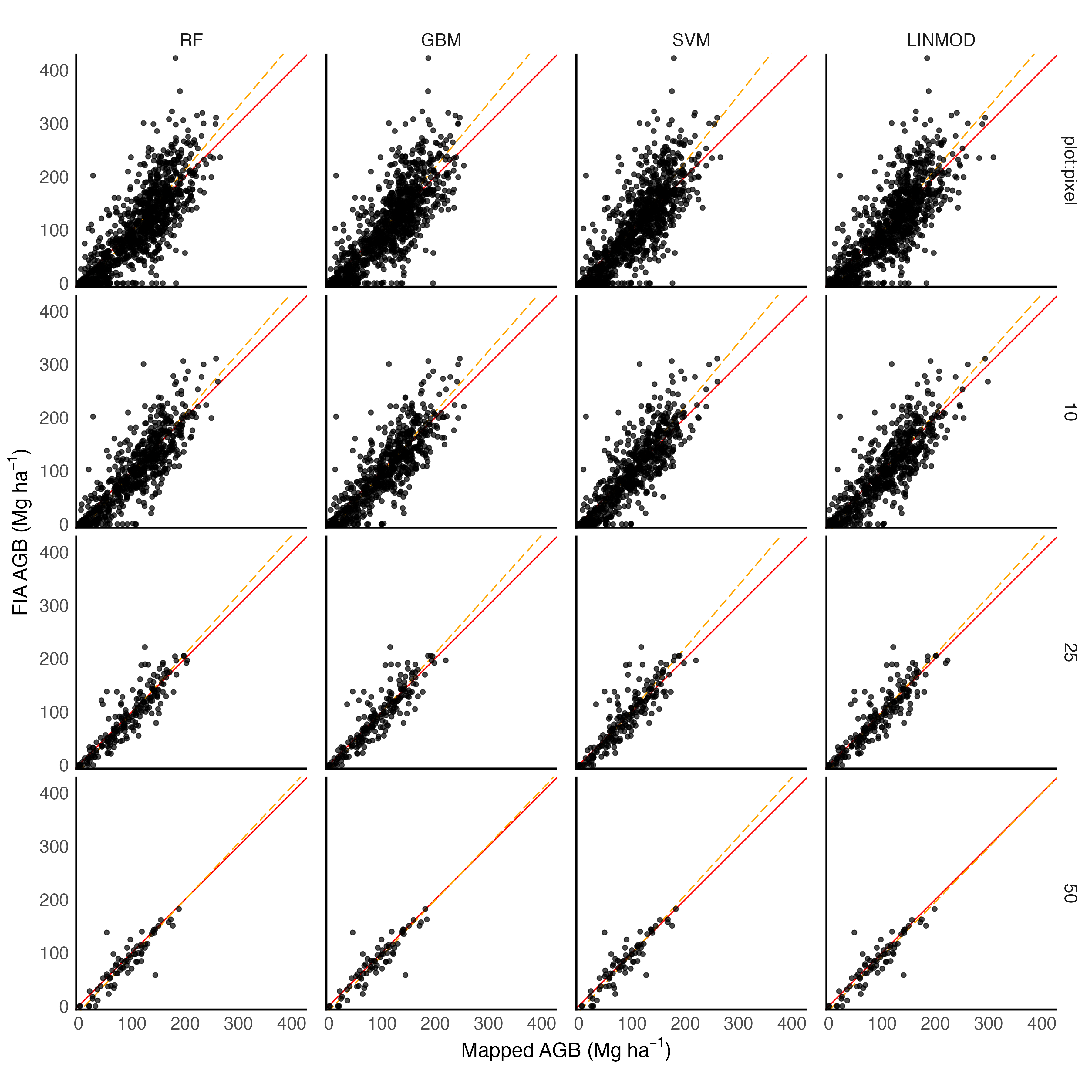} \caption{Comparing mapped AGB for all models to FIA estimated AGB across selected scales represented by distances between hexagon centroids (plot:pixel, 10km, 25km, and 50km). AGB values in Mg ha$^{-1}$. Geometric Mean Functional Regression trend line shown with dashed (orange) line, and 1:1 line shown with solid (red) line.}\label{fig:suppriemannscatter}
\end{figure}

\begin{figure}
\includegraphics[width=1\linewidth]{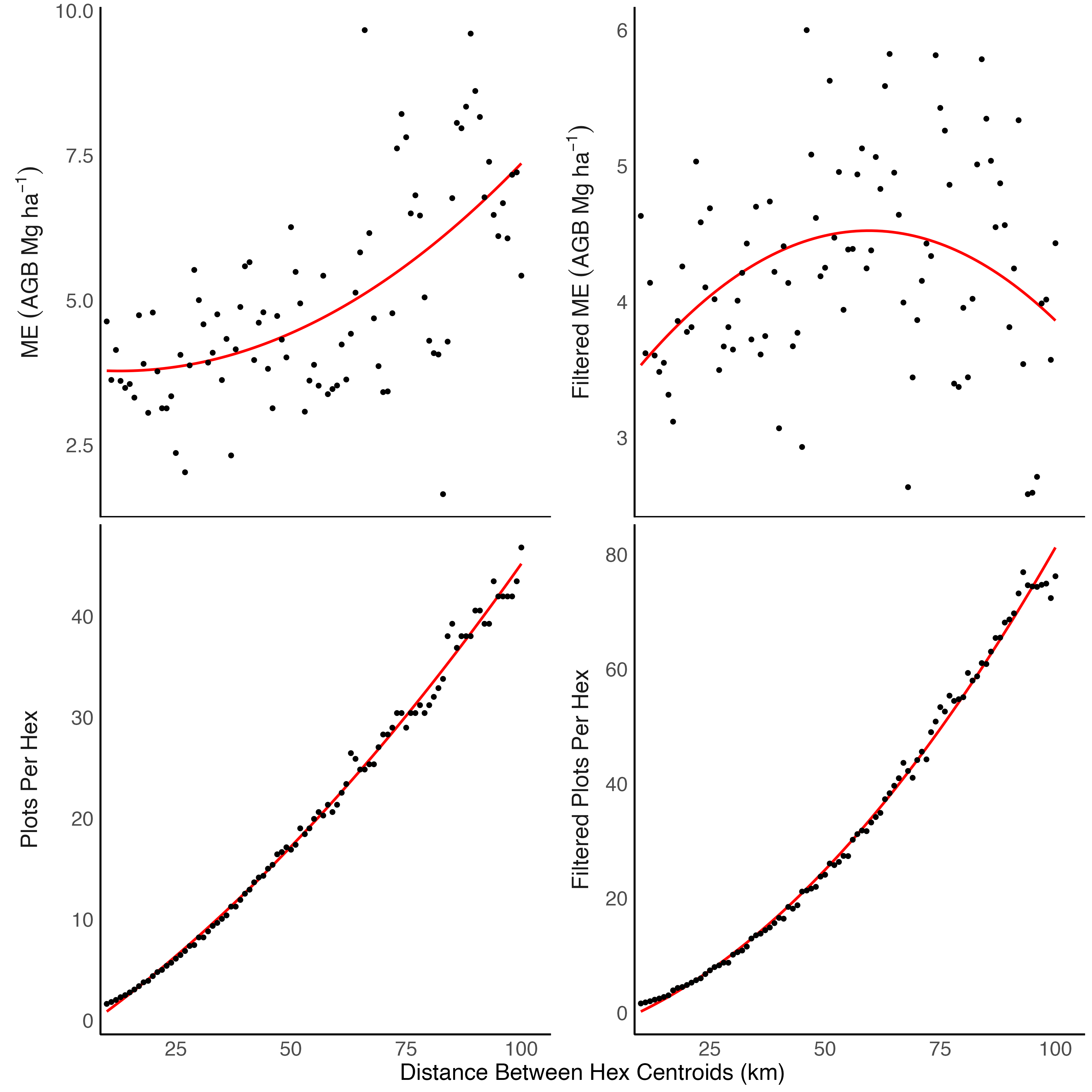} \caption{Mean error (as defined in Section 2.5) and plots per hexagon comparing mapped predictions to FIA estimates as a function of aggregation unit size (described by distances between hexagon centroids). Hexagons with an FIA plot density < 1 plot per 24,000 ha (1/10th the standard FIA sample intensity in NYS) were excluded. Red trend lines produced using quadratic regression.}\label{fig:filteredmeanerr}
\end{figure}

Agreement metrics and scatter plots for all component models are contained in
Table \ref{tab:suppacctable} and Figure \ref{fig:suppriemannscatter}
respectively.
To demonstrate how small samples within aggregation hexagons introduced noise
to estimates of ME,
we filtered hexagons where the density of FIA plots was \textless{} 1 plot per
24,000 ha (1/10 the FIA sampling density in NYS; Gray et al. (2012)), and plotted ME and
plots per hexagons for both the filtered set of hexagons as well as the full
set of hexagons separately (Figure \ref{fig:filteredmeanerr}).

\clearpage{}

\begin{table}

\caption{\label{tab:suppacctable}Map agreement results for select scales (in ha) and all models. Dist = distance between hex centroids in km; PPH = plots per hex; 
      n = number of comparison units (plots or hexagons);
      RMSE, MAE, ME in Mgha$^{-1}$. All metrics as defined in Section 2.5. Standard errors in parentheses, with a minimum of 0.01 for display.}
\centering
\fontsize{8}{10}\selectfont
\begin{tabular}[t]{lrrrlrrrrl}
\toprule
\multicolumn{1}{c}{Scale} & \multicolumn{1}{c}{Dist} & \multicolumn{1}{c}{n} & \multicolumn{1}{c}{PPH} & \multicolumn{1}{c}{Model} & \multicolumn{1}{c}{\% RMSE} & \multicolumn{1}{c}{RMSE} & \multicolumn{1}{c}{MAE} & \multicolumn{1}{c}{ME} & \multicolumn{1}{c}{R$^2$}\\
\midrule
 &  &  &  & RF & 44.88 & 40.94 (0.04) & 28.31 (0.85) & 4.96 (1.17) & 0.73 (0.01)\\

 &  &  &  & GBM & 45.12 & 41.16 (0.04) & 28.39 (0.85) & 3.44 (1.18) & 0.73 (0.01)\\

 &  &  &  & SVM & 45.37 & 41.39 (0.04) & 28.26 (0.87) & 0.64 (1.19) & 0.73 (0.01)\\

\multirow[t]{-4}{*}{\raggedright\arraybackslash Plot Pixel} & \multirow[t]{-4}{*}{\raggedleft\arraybackslash } & \multirow[t]{-4}{*}{\raggedleft\arraybackslash 1217} & \multirow[t]{-4}{*}{\raggedleft\arraybackslash } & LINMOD & 44.87 & 40.93 (0.04) & 28.12 (0.85) & 4.40 (1.17) & 0.73 (0.01)\\
\midrule
\addlinespace
 &  &  &  & RF & 37.93 & 34.04 (0.05) & 24.01 (0.89) & 5.18 (1.24) & 0.77 (0.01)\\

 &  &  &  & GBM & 38.52 & 34.57 (0.05) & 24.34 (0.90) & 3.67 (1.27) & 0.76 (0.01)\\

 &  &  &  & SVM & 38.18 & 34.27 (0.05) & 23.98 (0.90) & 0.82 (1.26) & 0.76 (0.01)\\

\multirow[t]{-4}{*}{\raggedright\arraybackslash 8660} & \multirow[t]{-4}{*}{\raggedleft\arraybackslash 10} & \multirow[t]{-4}{*}{\raggedleft\arraybackslash 739} & \multirow[t]{-4}{*}{\raggedleft\arraybackslash 1.65} & LINMOD & 37.95 & 34.06 (0.05) & 23.95 (0.89) & 4.63 (1.24) & 0.77 (0.01)\\
\midrule
\addlinespace
 &  &  &  & RF & 27.33 & 24.13 (0.13) & 17.02 (1.22) & 3.00 (1.70) & 0.81 (0.01)\\

 &  &  &  & GBM & 28.11 & 24.82 (0.14) & 16.99 (1.29) & 1.73 (1.76) & 0.80 (0.01)\\

 &  &  &  & SVM & 29.17 & 25.75 (0.15) & 17.70 (1.33) & -1.07 (1.83) & 0.78 (0.01)\\

\multirow[t]{-4}{*}{\raggedright\arraybackslash 54126} & \multirow[t]{-4}{*}{\raggedleft\arraybackslash 25} & \multirow[t]{-4}{*}{\raggedleft\arraybackslash 199} & \multirow[t]{-4}{*}{\raggedleft\arraybackslash 6.12} & LINMOD & 27.95 & 24.68 (0.13) & 17.13 (1.26) & 2.36 (1.75) & 0.80 (0.01)\\
\midrule
\addlinespace
 &  &  &  & RF & 25.79 & 21.11 (0.41) & 14.41 (1.83) & 7.03 (2.36) & 0.78 (0.01)\\

 &  &  &  & GBM & 26.04 & 21.31 (0.42) & 14.02 (1.90) & 5.73 (2.44) & 0.78 (0.01)\\

 &  &  &  & SVM & 25.07 & 20.52 (0.37) & 13.94 (1.79) & 3.11 (2.41) & 0.79 (0.01)\\

\multirow[t]{-4}{*}{\raggedright\arraybackslash 216506} & \multirow[t]{-4}{*}{\raggedleft\arraybackslash 50} & \multirow[t]{-4}{*}{\raggedleft\arraybackslash 72} & \multirow[t]{-4}{*}{\raggedleft\arraybackslash 16.9} & LINMOD & 25.86 & 21.17 (0.41) & 14.26 (1.86) & 6.26 (2.40) & 0.78 (0.01)\\
\midrule
\addlinespace
 &  &  &  & RF & 22.33 & 17.43 (0.78) & 11.63 (2.60) & 6.21 (3.26) & 0.74 (0.03)\\

 &  &  &  & GBM & 21.31 & 16.63 (0.68) & 11.13 (2.47) & 4.86 (3.18) & 0.76 (0.03)\\

 &  &  &  & SVM & 19.70 & 15.37 (0.57) & 10.42 (2.26) & 2.95 (3.02) & 0.80 (0.02)\\

\multirow[t]{-4}{*}{\raggedright\arraybackslash 866025} & \multirow[t]{-4}{*}{\raggedleft\arraybackslash 100} & \multirow[t]{-4}{*}{\raggedleft\arraybackslash 26} & \multirow[t]{-4}{*}{\raggedleft\arraybackslash 46.81} & LINMOD & 22.28 & 17.39 (0.71) & 11.58 (2.59) & 5.42 (3.30) & 0.74 (0.03)\\
\bottomrule
\end{tabular}
\end{table}

\clearpage

\hypertarget{s6-spatial-autocorrelation-of-mapped-residuals}{%
\section{S6: Spatial Autocorrelation of Mapped Residuals}\label{s6-spatial-autocorrelation-of-mapped-residuals}}

To provide context under conditions of spatial randomness
for the computed Moran's I statistics, 1000 simulations were produced for each
search radius.
In each simulation, the spatial structure of the observations is maintained,
but the associated residual information is shuffled randomly.
The upper 97.5th percentile and the lower 2.5th percentile of simulated
Moran's I statistics are selected to produce upper and lower envelopes
for the true Moran's I estimates.

\begin{figure}
\includegraphics[width=1\linewidth]{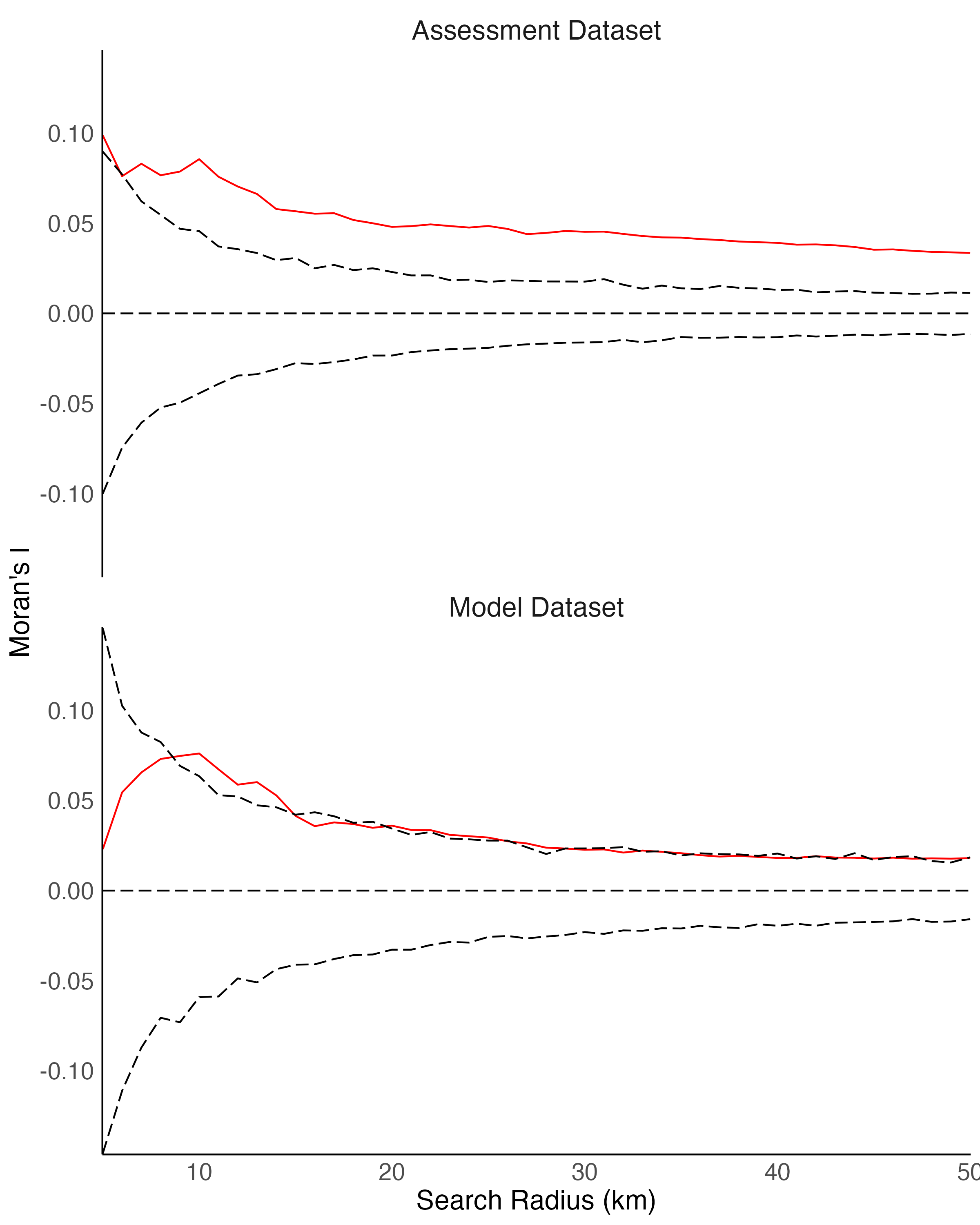} \caption{Global spatial autocorrelation (Moran's I) of the linear ensemble model residuals as a function of search radius for the assessment dataset and the model dataset (described in Section 2.2). Black dashed upper and lower envelopes represent the 95\% interval computed from 1000 bootstrap iterations of randomly re-assigning plot locations.}\label{fig:unnamed-chunk-2}
\end{figure}

\newpage{}

\hypertarget{references}{%
\section*{References}\label{references}}
\addcontentsline{toc}{section}{References}

\hypertarget{refs}{}
\begin{CSLReferences}{1}{0}
\leavevmode\vadjust pre{\hypertarget{ref-wwe}{}}%
Atlantic Inc. 2015. {``{NY\_WarrenWashingtonEssex\_Spring2015}.''} \url{ftp://ftp.gis.ny.gov/elevation/LIDAR/}.

\leavevmode\vadjust pre{\hypertarget{ref-mo}{}}%
Axis GeoSpatial, LLC. 2016a. {``{Axis GeoSpatial, LLC New York} Tiled {LiDAR}.''} \url{ftp://ftp.gis.ny.gov/elevation/LIDAR/}.

\leavevmode\vadjust pre{\hypertarget{ref-cr}{}}%
---------. 2016b. {``{New York Office of Information Technology Services} Classified {LiDAR} Tiles.''} \url{ftp://ftp.gis.ny.gov/elevation/LIDAR/}.

\leavevmode\vadjust pre{\hypertarget{ref-Breiman2001}{}}%
Breiman, Leo. 2001. {``Random Forests.''} \emph{Machine Learning} 45 (1): 5--32. \url{https://doi.org/10.1023/A:1010933404324}.

\leavevmode\vadjust pre{\hypertarget{ref-Draper1998}{}}%
Draper, Norman R, and Harry Smith. 1998. \emph{Applied Regression Analysis}. Vol. 326. John Wiley \& Sons.

\leavevmode\vadjust pre{\hypertarget{ref-Gray2012}{}}%
Gray, Andrew N, Thomas J Brandeis, John D Shaw, William H McWilliams, and Patrick Miles. 2012. {``{Forest Inventory and Analysis Database of the United States of America (FIA)}.''} \emph{Biodiversity and Ecology} 4: 225--31. \url{https://doi.org/10.7809/b-e.00079}.

\leavevmode\vadjust pre{\hypertarget{ref-Hosking1990}{}}%
Hosking, J. R. M. 1990. {``{L-Moments: Analysis and Estimation of Distributions Using Linear Combinations of Order Statistics}.''} \emph{Journal of the Royal Statistical Society. Series B (Methodological)} 52 (1): 105--24. \url{https://doi.org/10.1111/j.2517-6161.1990.tb01775.x}.

\leavevmode\vadjust pre{\hypertarget{ref-terrainr}{}}%
Mahoney, Michael. 2021. \emph{{{terrainr}: Landscape Visualizations in R and Unity}}. https://doi.org/\url{https://doi.org/10.5281/zenodo.5142763}.

\leavevmode\vadjust pre{\hypertarget{ref-Meyer2021}{}}%
Meyer, Hanna, and Edzer Pebesma. 2021. {``{Predicting into unknown space? Estimating the area of applicability of spatial prediction models}.''} \emph{Methods in Ecology and Evolution} 12 (9): 1620--33. \url{https://doi.org/10.1111/2041-210x.13650}.

\leavevmode\vadjust pre{\hypertarget{ref-as}{}}%
New York Office of Information Technology Services. 2016. {``{Allegany and Steuben Counties, New York Lidar}; Overall Project Metadata.''} \url{ftp://ftp.gis.ny.gov/elevation/LIDAR/}.

\leavevmode\vadjust pre{\hypertarget{ref-sw}{}}%
---------. 2017. {``Southwest 17 - Spring, {New York Lidar}; Classified Point Cloud.''} \url{ftp://ftp.gis.ny.gov/elevation/LIDAR/}.

\leavevmode\vadjust pre{\hypertarget{ref-co}{}}%
---------. 2018a. {``{LIDAR} Collection {(Ql2)} for {Cayuga} County and Most of {Oswego} County, {New York Lidar}; Classified Point Cloud.''} \url{ftp://ftp.gis.ny.gov/elevation/LIDAR/}.

\leavevmode\vadjust pre{\hypertarget{ref-swb}{}}%
---------. 2018b. {``Southwest 17-b - Fall, {New York Lidar}; Classified Point Cloud.''} \url{ftp://ftp.gis.ny.gov/elevation/LIDAR/}.

\leavevmode\vadjust pre{\hypertarget{ref-egl}{}}%
---------. 2019. {``{LIDAR} Collection {(QL2) for Erie, Genesee, and Livingston} Counties {New York Lidar}; Classified Point Cloud.''} \url{ftp://ftp.gis.ny.gov/elevation/LIDAR/}.

\leavevmode\vadjust pre{\hypertarget{ref-tax}{}}%
NYSGPO. 2018. \emph{Property Type Classification Codes}. New York State Department of Taxation; Finance. \url{https://www.tax.ny.gov/research/property/assess/manuals/prclas.htm}.

\leavevmode\vadjust pre{\hypertarget{ref-cef}{}}%
Quantum Spatial. 2016. {``{Clinton-Essex-Lake Champlain New York 2015 LiDAR USGS} Contract: {G10pc00026} {Task} Order Number: G10pc00026 and G14pd000943 (Modification) {NY\_ClintonEssex\_2015}.''} \url{ftp://ftp.gis.ny.gov/elevation/LIDAR/}.

\leavevmode\vadjust pre{\hypertarget{ref-os}{}}%
Quantum Spatial, Inc. 2017a. {``{New York FEMA 2016 Ql2 LiDAR - Central} Zone {AOI}; Classified Point Cloud.''} \url{ftp://ftp.gis.ny.gov/elevation/LIDAR/}.

\leavevmode\vadjust pre{\hypertarget{ref-fsl}{}}%
---------. 2017b. {``{New York FEMA 2016 Ql2 LiDAR - East} Zone {AOI}; Classified Point Cloud.''} \url{ftp://ftp.gis.ny.gov/elevation/LIDAR/}.

\leavevmode\vadjust pre{\hypertarget{ref-fshf}{}}%
---------. 2018. {``Lidar Collection {(Ql2)} of All or Part of {Schuyler, Seneca, Steuben, Tompkins, Wayne and Yates} Counties, {NY Lidar}; Classified Point Cloud.''} \url{ftp://ftp.gis.ny.gov/elevation/LIDAR/}.

\leavevmode\vadjust pre{\hypertarget{ref-Ricker1984}{}}%
Ricker, WE. 1984. {``Computation and Uses of Central Trend Lines.''} \emph{Canadian Journal of Zoology} 62 (10): 1897--1905.

\leavevmode\vadjust pre{\hypertarget{ref-Riemann2010}{}}%
Riemann, Rachel, Barry Tyler Wilson, Andrew Lister, and Sarah Parks. 2010. {``{An effective assessment protocol for continuous geospatial datasets of forest characteristics using {USFS} Forest Inventory and Analysis ({FIA}) data}.''} \emph{Remote Sensing of Environment} 114 (10): 2337--52. \url{https://doi.org/10.1016/j.rse.2010.05.010}.

\leavevmode\vadjust pre{\hypertarget{ref-fasterize}{}}%
Ross, Noam. 2020. \emph{{fasterize: Fast Polygon to Raster Conversion}}. \url{https://CRAN.R-project.org/package=fasterize}.

\leavevmode\vadjust pre{\hypertarget{ref-3c}{}}%
United States Geological Survey. 2015a. {``{LAS}.''} \url{ftp://ftp.gis.ny.gov/elevation/LIDAR/}.

\leavevmode\vadjust pre{\hypertarget{ref-sco}{}}%
---------. 2015b. {``{LAS}.''} \url{ftp://ftp.gis.ny.gov/elevation/LIDAR/}.

\leavevmode\vadjust pre{\hypertarget{ref-gl}{}}%
---------. 2015c. {``{LAS} Extents.''} \url{ftp://ftp.gis.ny.gov/elevation/LIDAR/}.

\leavevmode\vadjust pre{\hypertarget{ref-li}{}}%
Woolpert, Inc. 2014a. {``{USGS Long Island New York Sandy Lidar} Classified {LAS} 1.2.''} \url{ftp://ftp.gis.ny.gov/elevation/LIDAR/}.

\leavevmode\vadjust pre{\hypertarget{ref-nyc}{}}%
---------. 2014b. {``{USGS New York CMGP Sandy Lidar}.''} \url{ftp://ftp.gis.ny.gov/elevation/LIDAR/}.

\end{CSLReferences}